\documentclass[pdflatex,sn-mathphys]{sn-jnl}

\jyear{2021}%

\theoremstyle{thmstyleone}%
\theoremstyle{thmstyletwo}%

\theoremstyle{thmstylethree}%

\begin{document}

\title[Article Title]{Study of the front-end signal for the 3-inch PMTs instrumentation in JUNO}


%

\author[1,2]{Diru Wu}
\author*[1]{Jilei Xu}
\author[1]{Miao He}
\author[1]{Zhimin Wang}
\author[1,2]{Ziliang Chu}

\affil[1]{Institute of High Energy Physics, \city{Beijing}, \postcode{100049}, \country{China}}
\affil[2]{University of Chinese Academy of Sciences, \city{Beijing}, \postcode{100049}, \country{China}}

%
%
%
%
%


\abstract{25,600 3-inch PMTs will be installed in Jiangmen Underground Neutrino Observatory (JUNO) to achieve more precise energy calibration and to extend the physics detection potential. Performances of all bare PMTs have been characterized and these PMTs are being instrumented with the high voltage divider, underwater front-end cable, and connector. In this paper, we present a dedicated study on signal quality at different stages of the instrumentation. An optimized high voltage ratio was confirmed and finalized which improved the PMT transit time spread by 25\%. The signal charge was attenuated by 22.5\% (13.0\%) in the 10~m (5~m) cable and it required the addition of 45~V (23~V) to compensate for the loss of PMT gain. There was a 1\% overshoot following the PMT signal and no sign of reflection in the connector. A group of 16 3-inch PMTs with the full instrumentation was installed in the JUNO prototype detector together with a few 8-inch and 20-inch PMTs, which showed good stability and demonstrated a photon detection system with multiple types of PMTs.}




\maketitle

\section{Introduction}
\label{sec.intro}
The Jiangmen Underground Neutrino Observatory (JUNO) is proposed to determine the neutrino mass ordering by a 20~kiloton liquid scintillator (LS) detector, located $\sim$700~m underground in Guangdong province, China \cite{juno-yellowbook, juno-CDR}. The LS detector is contained in an acrylic spherical shell with an inner diameter of 35.4~m and a thickness of 12~cm. To reach a sensitivity better than 3 standard deviations of the neutrino mass ordering after 6 years of data taking, the energy scale uncertainty should be controlled within $1\%$ and the effective energy resolution should be better than $3\%$ at 1~MeV. This requires 17,612 20-inch PMTs (Large PMTs or LPMTs) and 25,600 3-inch PMTs (Small PMTs or SPMTs) closely packed around the acrylic sphere. SPMTs will be installed between the gaps of LPMTs to make a double calorimetry system, which can help the LPMT system to improve physics capabilities by reducing the systematic uncertainties related to the energy measurement \cite{doubleCal-xujl}.

\begin{figure}[H]
\centering
\includegraphics[width=0.6\textwidth]{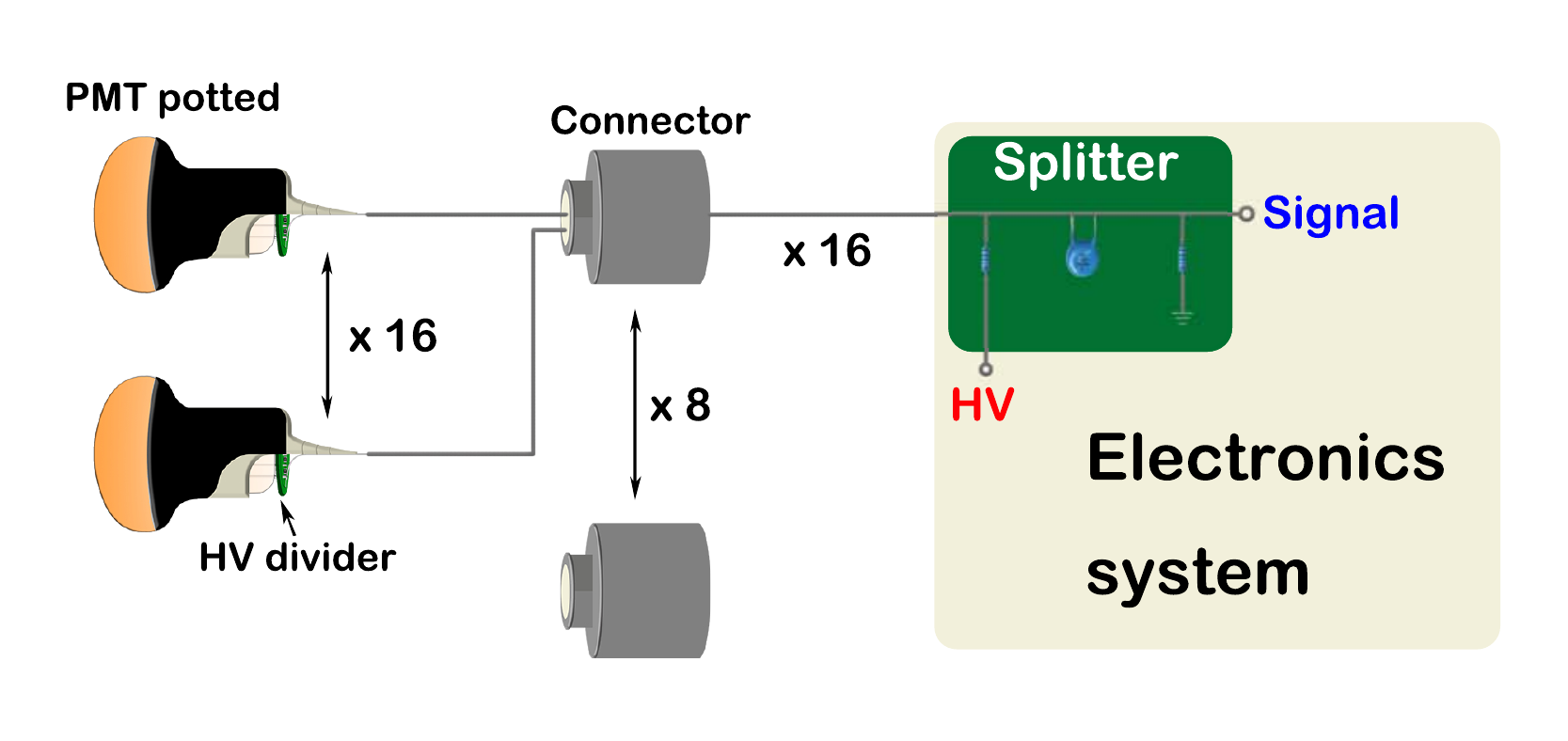}
\caption{Conceptual diagram of SPMT system in JUNO. Each PMT is potted with a waterproof shell and has only one cable for HV supply and signal output. Every 16 PMTs are linked to one connector. 8 connectors are mounted on one UWB. Splitters are used to decouple the signal and HV power.}
\label{TotalSys}
\end{figure}

All the PMTs are immersed in a cylindrical pool with both diameter and height 43.5~m filled with ultra-pure water \cite{juno-PPNP}. The SPMT system is an independent photon detection system consisting of not only PMTs but also the front-end instrumentation and readout electronics (\cite{CATIROC}, Fig.~\ref{TotalSys}), all of which are underwater. The PMT is soldered on a high voltage (HV) divider, and the signal, as well as the HV, are transmitted through a customized RG178 coaxial cable with an HDPE jacket. The PMT pins and the divider is sealed in an ABS plastic shell for underwater proofing. 16 PMTs with the same cable length, either 5~m or 10~m, are sealed in a multi-channel connector. The connector consists of a plug attached to the PMT, and a receptacle sitting on the wall of a stainless steel box as the container of the readout electronics. When the plug and the receptacle are 
matched, the connection is waterproof thus ensuring the full system working underwater. Both the cable and the underwater connector were designed and produced by AXON' Interconnect Co., Ltd \cite{axon} according to the requirement of JUNO.

All the 3-inch PMTs are dynode PMTs of type XP72B22 which were produced by the HZC company~\cite{HZC-company}, and their performances have been characterized~\cite{mass-production}. In the final configuration of JUNO layout, the SPMT signal is also influenced by the design of the HV divider, the cable and the connector. In this paper, we will present the analog signal study with the PMT instrumentation up to the connector in Sec.~\ref{sec.lab}, including optimization of the voltage ratio in the HV divider, the attenuation in the front-end cable, and the reflection in the connector. A group of 16 3-inch PMTs was installed in a JUNO detector prototype and the performances will be displayed in Sec.~\ref{section:prototype} and compared with 8-inch and 20-inch PMTs in the same detector. The conclusion is presented in Sec.~\ref{sec.summary}.

\section{The optimization of front-end part}
\label{sec.lab}

\subsection{Measurement system}
\label{section:testsystem}
A measurement system was built to collect either waveform or charge information of PMT signals. 
In this system, a 405~nm laser was used as a light source; the light beam was driven through an optical fiber into a dark box where a lens creates a light spot with a diameter of 80~mm to properly cover the surface of SPMT photocathode. A signal generator provided two sync pulses, one used to trigger the laser driver and the other used to trigger a signal recorder (an oscilloscope, RIGOL DS6104 or a QDC, CAEN V965 or a Flash ADC, CAEN DT5751). A splitter was made to decouple HV and signals, and its Performance was studied in Ref.~\cite{luofj-overshoot}.

\begin{figure}[H]
    \centering
    \includegraphics[width=0.4\textwidth]{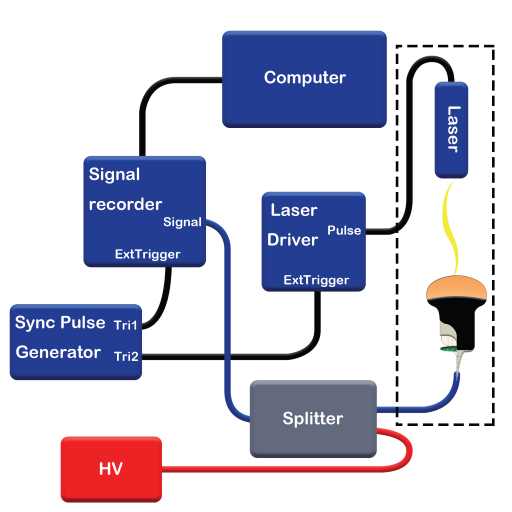}
    \caption{Diagram of the measurement system. The system was used to collect charge or waveform. The PMT and laser fiber were in a dark box.}
    \label{System}
\end{figure}

The laser light pulse intensity can be adjusted by the driven voltage to accommodate different studies. In particular, for the single photoelectron (SPE) measurement, the light intensity was turned down to about one SPE signal every 10 triggers, which means the probability that the PMT detected a laser signal was about 0.1. The number of photoelectrons collected by the first dynode obeys Poisson distribution:

\begin{equation}
    P(n;\mu)=\frac{e^{-\mu}\mu^{n}}{n!},
\end{equation}
where $n$ is the number of detected photoelectrons in each trigger with $\mu\approx0.1$ as the expectation. In this case, there will be $\sim95\%$ probability to have single PE (SPE) in all signals. 

\subsection{High Voltage ratio optimization}
\label{section:HVRatio-opt}
\begin{figure}[H]
    \centering
    \includegraphics[width=0.4\textwidth]{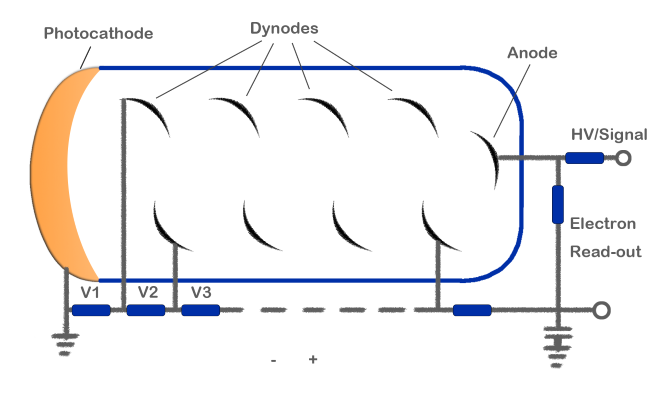} \\
    \caption{The high voltage divider and structure of dynode PMT. The blue rectangles represent resistors. V1, V2, V3 ... are the voltages on each resistor.} 
    \label{HVRatio}
\end{figure}

The divider was used to distribute the high voltage to the PMT dynodes by 11 series resistors, as shown in Fig.~\ref{HVRatio}. It is well known that the HVs of the first two dynodes play a more important role than the rest to collect the secondary electrons and to reduce the transit time spread (TTS). The HV ratio was recommended by HZC as 3:1:1...1, abbreviated as 3:1:1 since the rest of HV is the same as the third one. Using this HV ratio, the typical TTS of the JUNO SPMT is about 5~ns in terms of full width at half maxima (FWHM) for single PE signals, which means TTS ($\sigma$) \footnote{TTS ($\sigma$) was defined as FWHM / 2.36 in this paper. Some PMT factories and papers use FWHM only. For physics study, the $\sigma$ from Gaussian error is more widely used, which is about 2.36 times smaller than FWHM in mathematics.} is 2.1~ns. However, early studies~\cite{spmt-base, wg, caocy} in JUNO show that 3:2:1 would have lower TTS. In this paper, we used PMTs, cables, and connectors from the JUNO mass production to comprehensively study the SPMT's performances with 7 different HV ratios: 1:1:1, 2:1:1, 3:1:1, 4:1:1, 5:1:1, 3:2:1 and 3:3:1. TTS is the major concern in this study as it has a direct impact on the event vertex and track reconstruction in the detector, while the working HV, the SPE resolution, and the Peak-to-Valley (PV) ratio were also investigated at the same time.

\subsubsection{Gain calibration}
\label{section:calibration}
The working HVs at the nominal gain 3$\times$10$^{6}$ were firstly calibrated for different HV ratios. The PMT charge spectra were recorded by the QDC and were fitted with the PMT realistic response function which takes into account the ideal PMT's spectrum ($S_{ideal}(x')$) and background charge distribution ($B(x)$). 

\begin{equation}
\begin{aligned}
    S_{real}(x)&=S_{ideal}(x')\otimes B(x)\\
    &=\sum_{n=0}^{N}\frac{\mu^n e^{-\mu}}{n!}\left[(1-w)G_{n}(x-Q_{0})+w I_{G_{n}\otimes E}(x-Q_{0}) \right]
    \label{SReal}
\end{aligned}
\end{equation}

Where $G_n$ is a Gaussian distribution that is used to describe the charge of n photoelectrons signal.

\begin{equation}
    G_{n}=\frac{1}{\sigma_{1}\sqrt{2\pi n}}\,exp\left(-\frac{(x-nQ_{1})^{2}}{2n\sigma_{1}^{2}} \right)
    \label{nPEDis}
\end{equation}

\begin{figure}[H]
    \centering
    \includegraphics[width=0.45\textwidth]{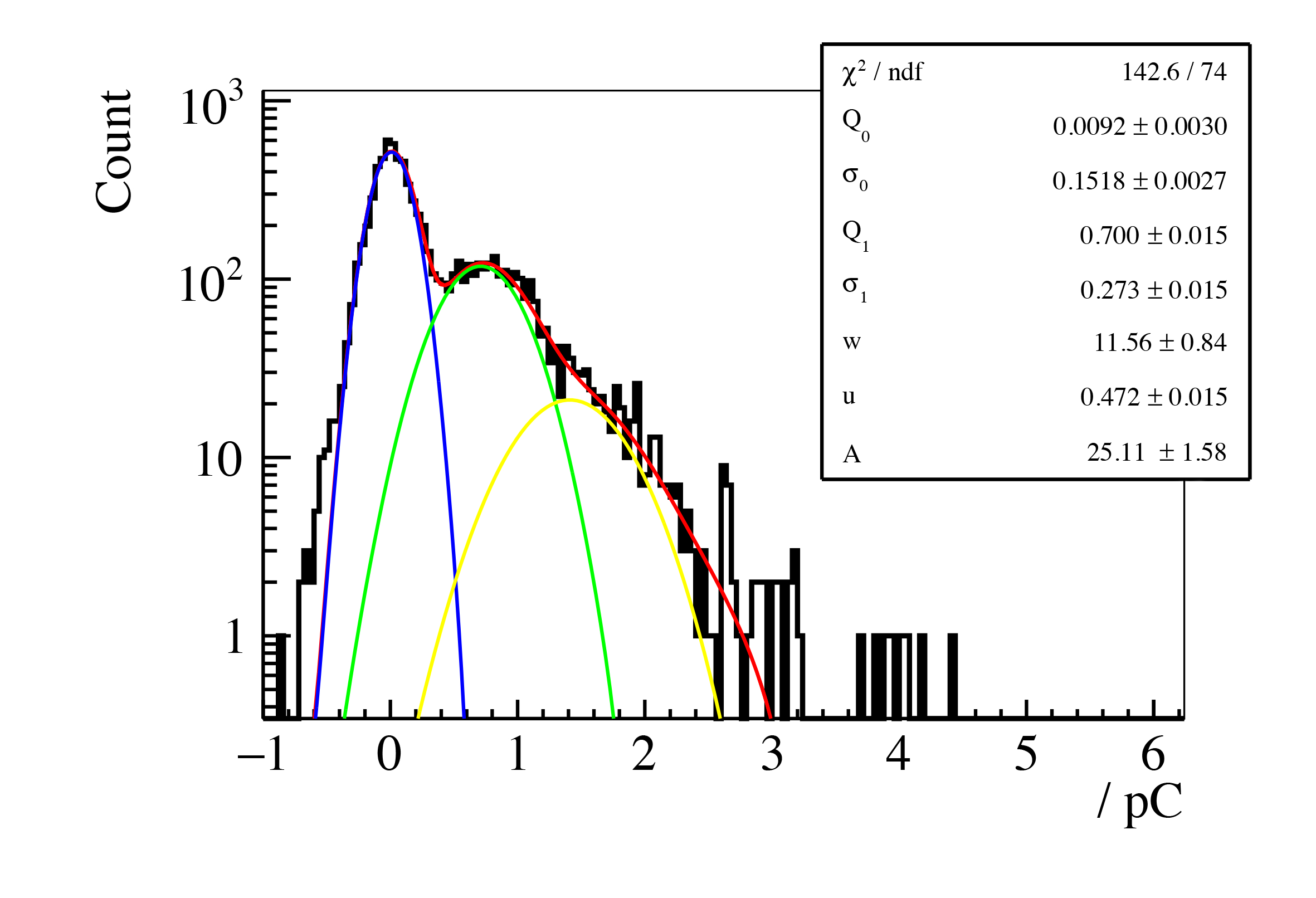}
    \caption{Measured charge spectrum overlapped with the fitting function in red. The green and yellow lines describe the single PE and double PE components, and the blue line is pedestal.}
    \label{SpectrumFit}
\end{figure}

More details of Eq.~\ref{SReal} can be found in Ref.~\cite{Calibration}. Where $N$ is the max number of PEs involved during a fitting process in the study, $Q_{0}$ is the pedestal position of the spectrum, $Q_{1}$ is the average charge of SPE signals, $\sigma_{1}$ is the standard deviation of SPE signals. $w$ is the probability that a measured signal is accompanied by a discrete processes noise. An example of a charge spectrum is shown in Fig.~\ref{SpectrumFit}, overlapped with the fitting results. The gain $G$ can be obtained by $G={Q_1}/{Q_e}$, where $Q_e$ is the charge of one electron. The HV was tuned till the difference of the fitted gain and the nominal gain was smaller than $0.1\times10^6$.

\begin{figure}[H]
    \centering
        \includegraphics[width=0.45\textwidth]{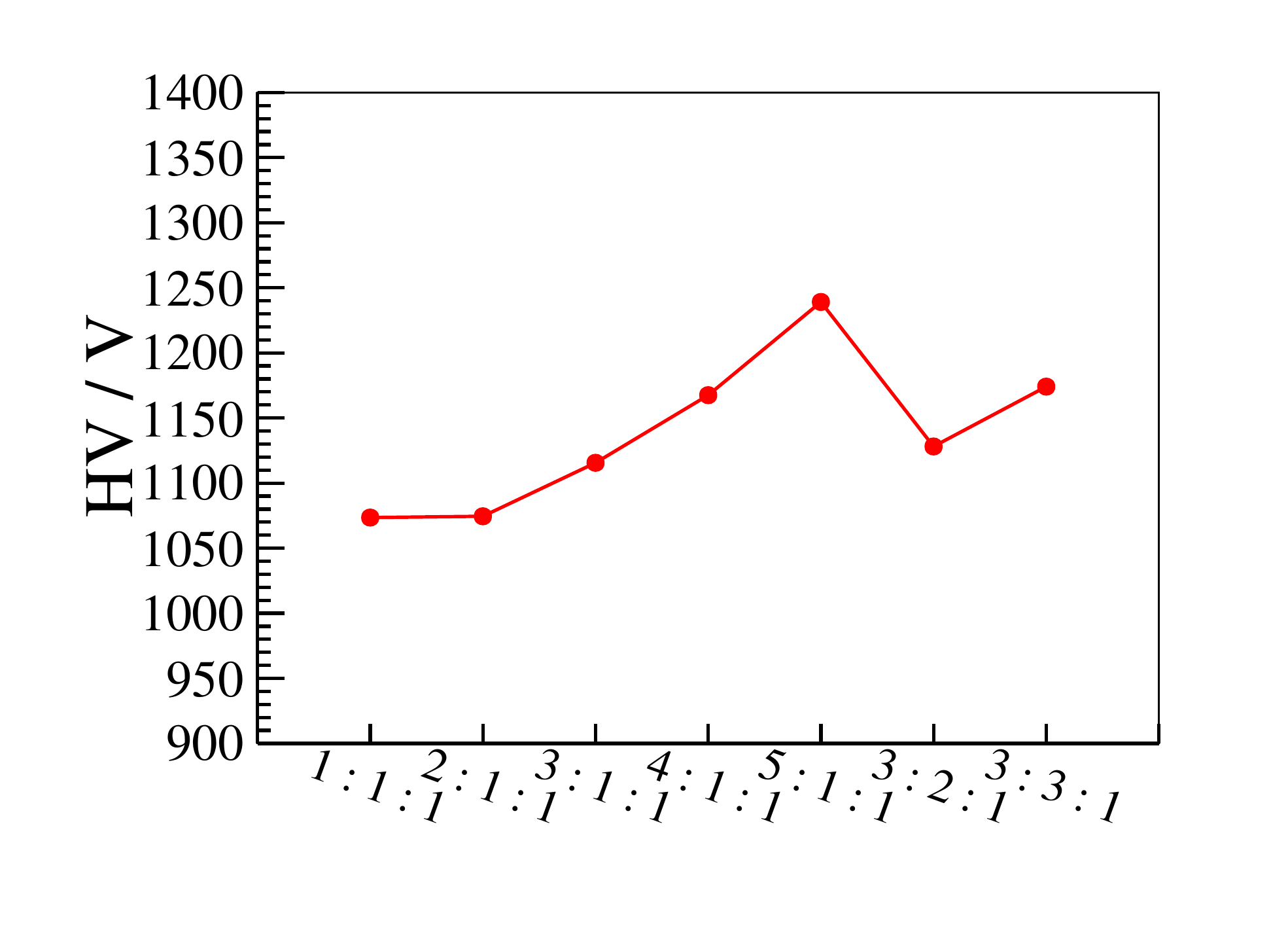}
    \caption{
    The averaged HVs for seven HV ratios normalized to the HV of 3:2:1 ratio.}
    \label{HVVsRatio}
\end{figure}

The HVs of Eight PMTs were calibrated with seven different HV ratios (Fig.~\ref{HVVsRatio}). Because different PMT has different nominal HV, the average voltage ($A$) of different PMTs at HV ratio 3:2:1 configuration was calculated. Under this configuration, the voltage of each PMT ($V_{i}$) had their own scale factor compared with $A$, $f_{i}=A/V_{i}$. For another configuration, each PMT's voltage can be scaled by a factor $f_{i}$, and the average voltage can be calculated. 

As the figure shows, a higher HV is needed to compensate for the loss of gain when the relative HV at the first or second dynode is enlarged. The maximum increase of HV is around 200~V for our investigated ratios.

\subsubsection{TTS measurement}
\label{section:TTS}

TTS($\sigma$) is expressed by $\sigma$ of the distribution of transit time which is the time between photoelectron emission and the moment when electrons are captured by the anode. The distribution is caused by the distribution of kinetic energy and direction of photoelectron generated by the photocathode \cite{linan}. The reactor neutrino vertex event can be reconstructed based on the relative arrival time from the interaction point to different PMTs, and PMT time resolution is contained in \cite{juno-yellowbook, juno-CDR}. The vertex reconstruction was largely influenced by TTS, the smaller the TTS, the better the vertex reconstruction.

In TTS measurement, the signal hit time can be get from PMT signal waveform fitted by a log-normal function. The simple fit function was inferred in paper \cite{WaveformModel}. For more convenient fitting, a constant value of baseline ($B$) and signal start time ($s$) were added to upgrade the fit function. The new function is

\begin{equation}
\begin{aligned}
    U(t)=A\,exp\left(-\frac{1}{2}\left(\frac{ln\left(\frac{t-s}{\tau - s} \right)}{\sigma} \right)^{2} \right)+B
    \label{WaveFit}
\end{aligned}
\end{equation}

\begin{figure}[H]
    \centering
    \subfigure[]{
        \includegraphics[width=0.45\textwidth]{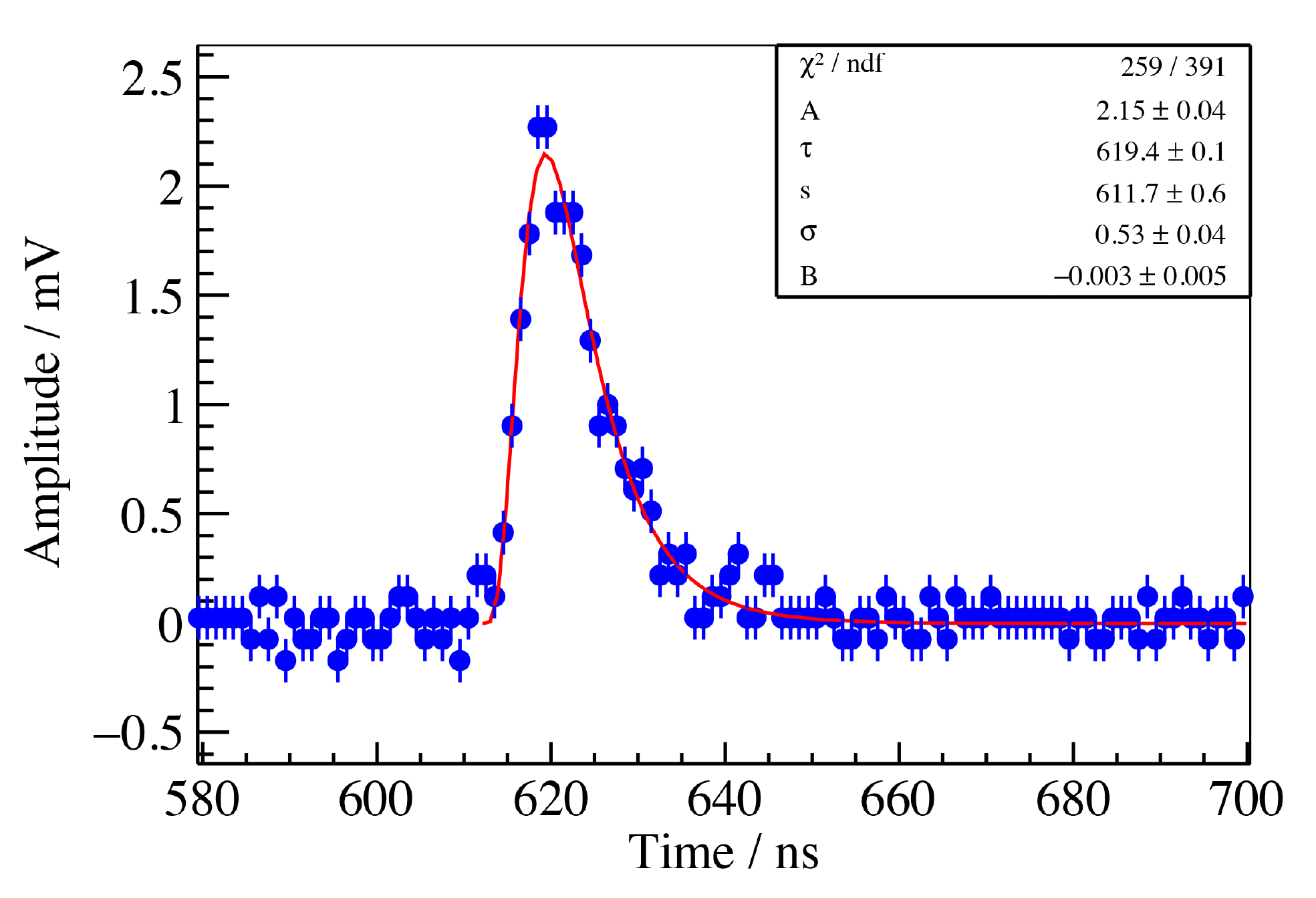}
    }
    \subfigure[]{
        \includegraphics[width=0.45\textwidth]{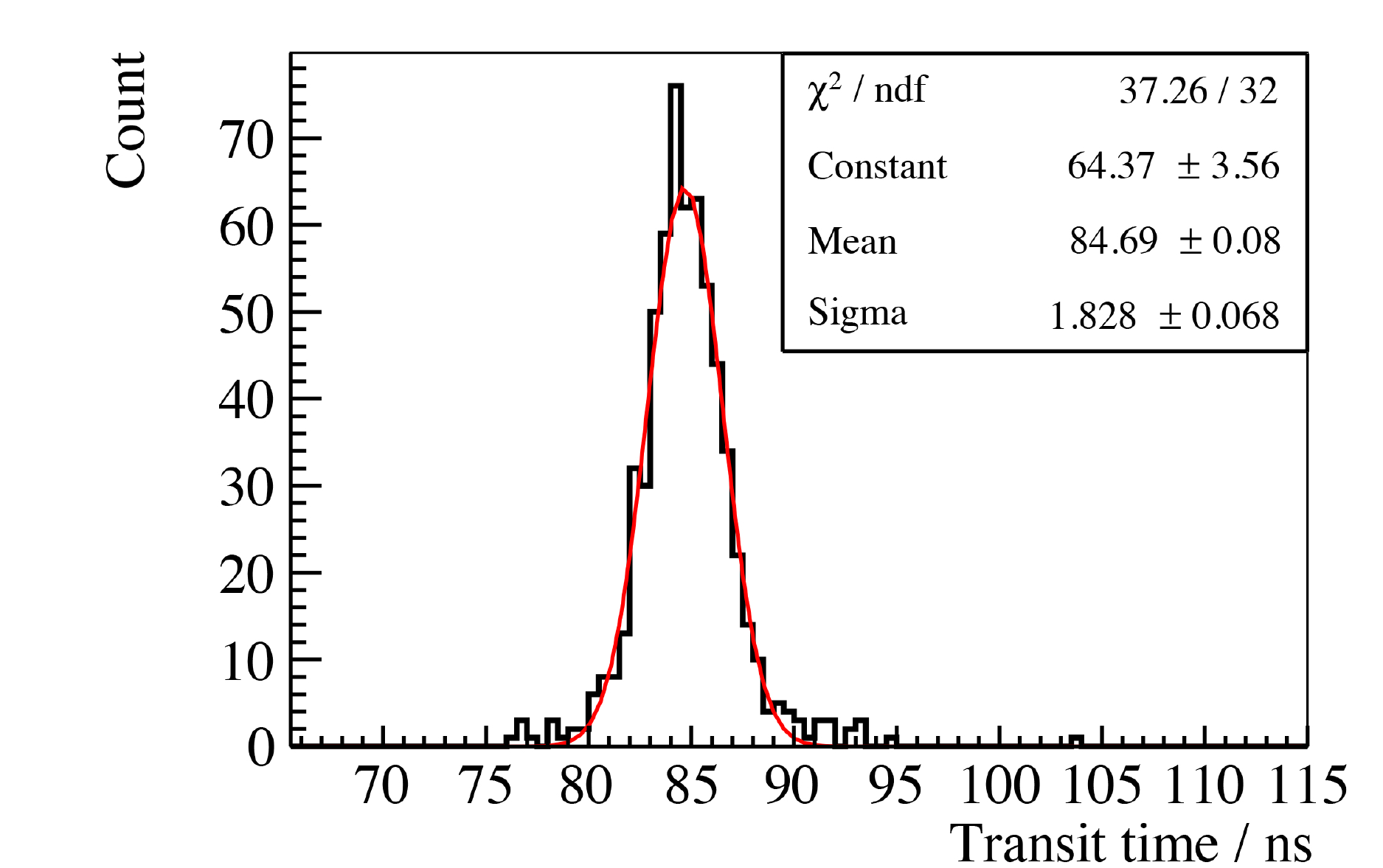}
    }
    \caption{(a) An example of SPE waveform fitted with the log-normal Eq.~\ref{WaveFit}. (b) Distribution of the transit time for one PMT fitted with the Gaussian function.
    }
    \label{TTSWave}
\end{figure}

Where $A$ is the amplitude of the pulse, $\sigma$ is a shape factor related to the width of the pulse,  $\tau$ is the peak position used as the signal hit time. The function fit result for a SPE waveform is shown in Figure~\ref{TTSWave}~(a). The fit can well describe the waveform of the signal and the hit time $\tau$ could be deduced. By subtracting the trigger time from the hit time, a measured PMT transit time distribution is shown in Fig.~\ref{TTSWave}~(b). TTS($\sigma$) is extracted by fitting the distribution of transit time with a Gaussian function (1.83 ns for the example PMT in the figure). The trigger time fluctuation can be relatively ignored since the time precision of the laser driver is at the picosecond level. 

\subsubsection{HV ratio Finalization}

\begin{figure}[H]
    \centering
    \subfigure[]{
        \includegraphics[width=0.3\textwidth]{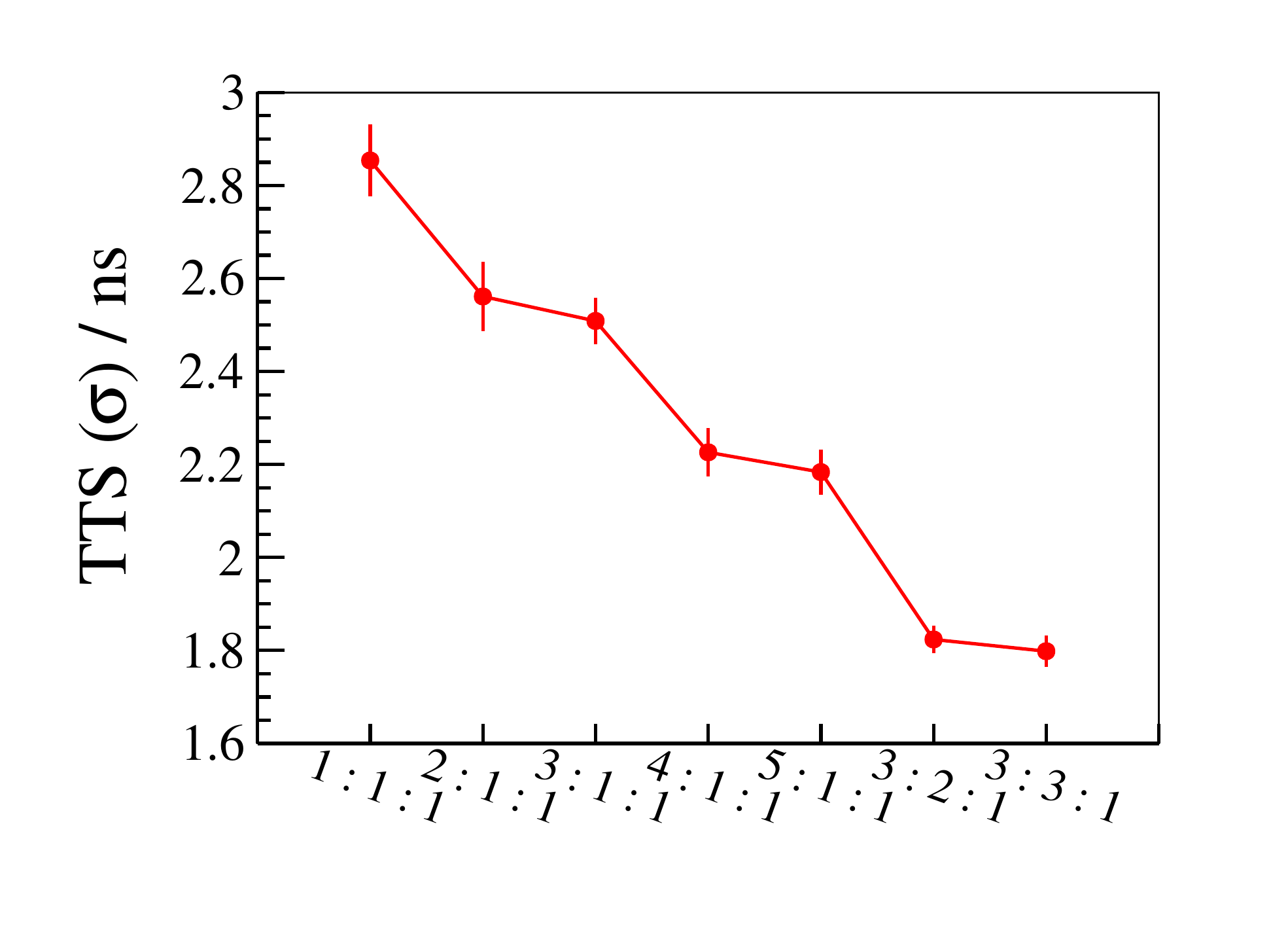}
    }
    \subfigure[]{
        \includegraphics[width=0.3\textwidth]{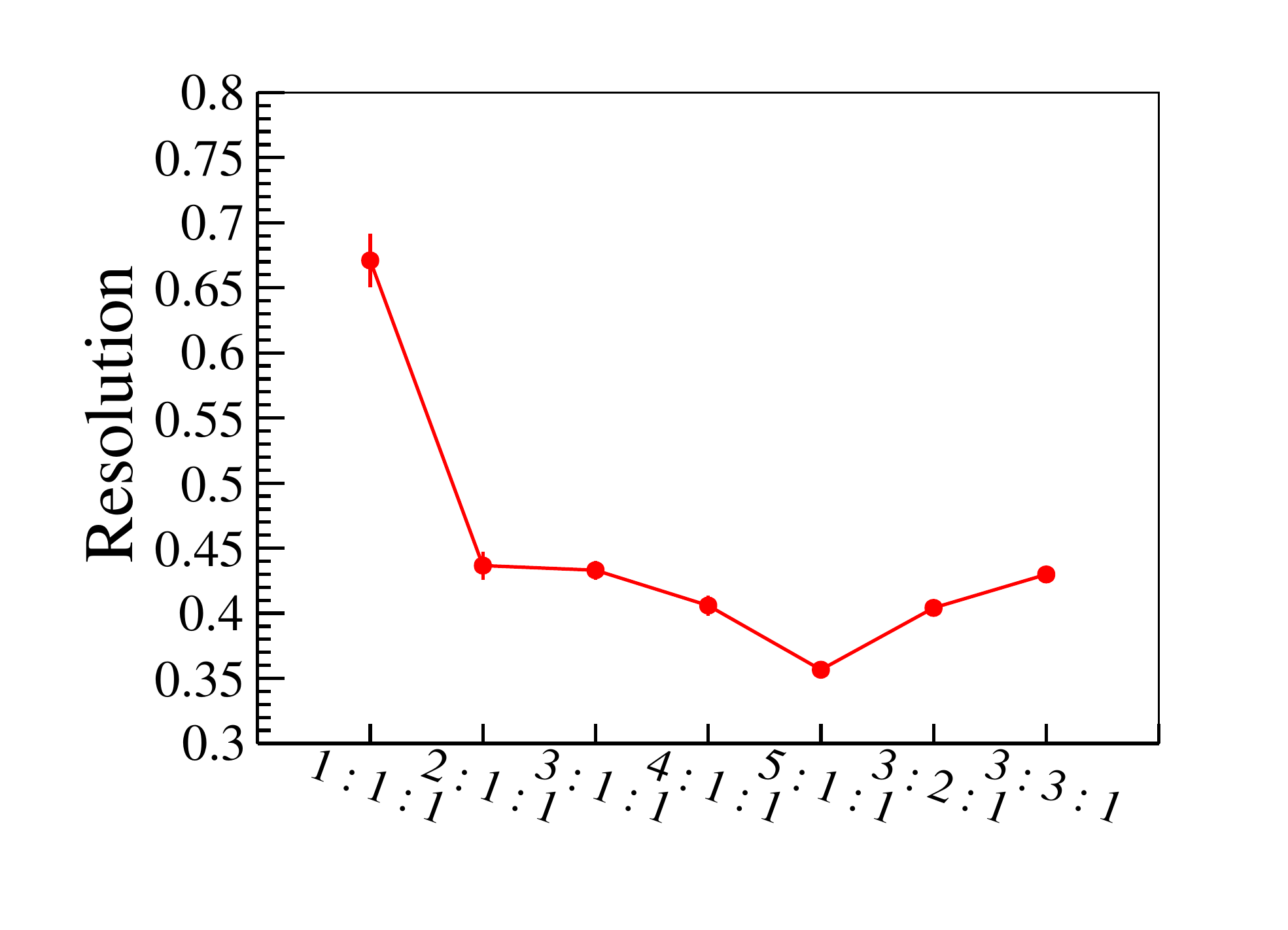}
    }
    \subfigure[]{
        \includegraphics[width=0.3\textwidth]{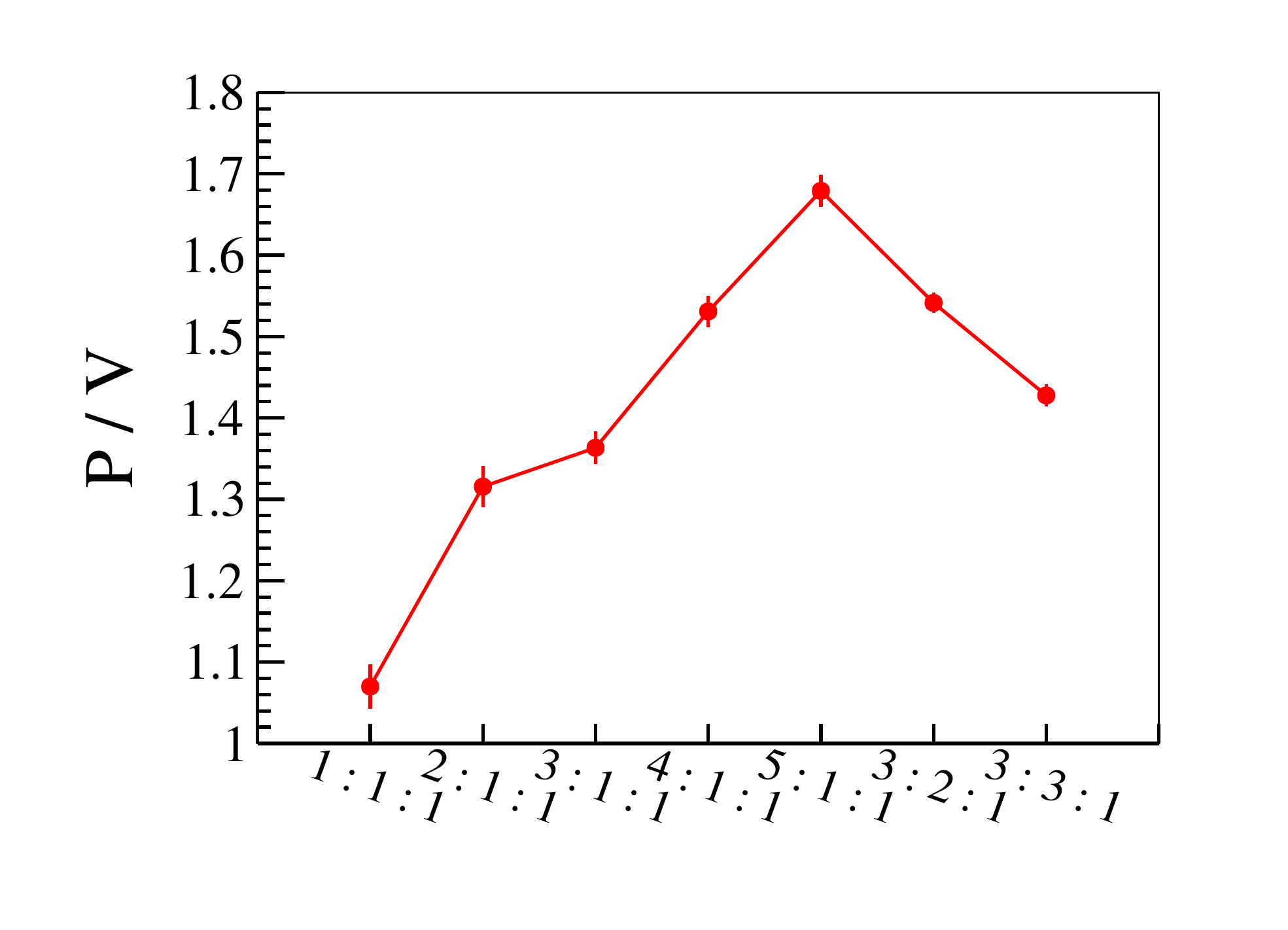}
    }
    \caption{The average TTS, SPE resolution, PV ratio of 8 PMTs under different HV ratios normalized to the average of HV ratio 3:2:1.}
    \label{TTSVsRatio}
\end{figure}

The average TTS versus HV ratio with the same normalization method in section \ref{section:calibration} was measured (Fig.~\ref{TTSVsRatio}~(a)). The HV ratios of 3:2:1 and 3:3:1 have better TTS than others with an improvement of $(68\pm6)\%$ and $(71\pm6)\%$ respectively compared to that of HV ratio 3:1:1 which is recommended by HZC.

For more precisely comparison, the SPE resolution ( $\sigma_{1}/Q_{1}$) and PV ratio are shown in Fig.~\ref{TTSVsRatio}~(b) and Fig.~\ref{TTSVsRatio}~(c). Except the ratio 1:1:1 which gets the worst resolution, all the others can meet JUNO's requirement (SPE resolution < 45\%) and the ratio 5:1:1 is the best one. The PV ratio of HV ratio 5:1:1 configuration is the best. The results of PV ratio and SPE resolution show that the HV between the cathode and the first dynode has a significant influence on the resolution. In the PMT, the distance between the cathode and the first dynode is the longest, so, the flight time of the photoelectron in this distance is most sensitive to the initial velocity of the electron. Higher voltage applied at this first stage can efficiently constrain the free motion of electrons with the overall benefit to decrease the transit time and increase the collection efficiency. 

Taking into account the importance of TTS to the event vertex reconstruction, 3:2:1 and 3:3:1 were considered as HV ratio candidates because of much smaller TTS than ratio of 3:1:1. For further comparison, the addition of 46~V is needed by the ratio 3:3:1 to reach the nominal gain compared to the ratio 3:2:1 (Fig.~\ref{HVVsRatio}). Therefore, 3:2:1 was selected as the final design for the voltage divider of JUNO SPMTs. This result is also consistent with the well-known conical HV ratio.


\subsection{Signal with longer cable}
\label{section:caleLength-opt}

All the bared 3-inch PMTs of the JUNO were characterized at the work stations of HZC with a front-end cable length less than 1~m. It is well known that the analog signal attenuates in the cable, therefore performances of PMT signals after the transmission in the JUNO cable have to be examined. In this study, we set up a test bench and measured several charge-related and time-related parameters using different cable lengths. An attenuation calculation method was developed to predict the signal shape with any cable length, which is useful for cable choosing in large-scale experiments.

\subsubsection{Theoretical calculation}
 
\begin{figure}[H]
    \centering
    \includegraphics[width=0.45\textwidth]{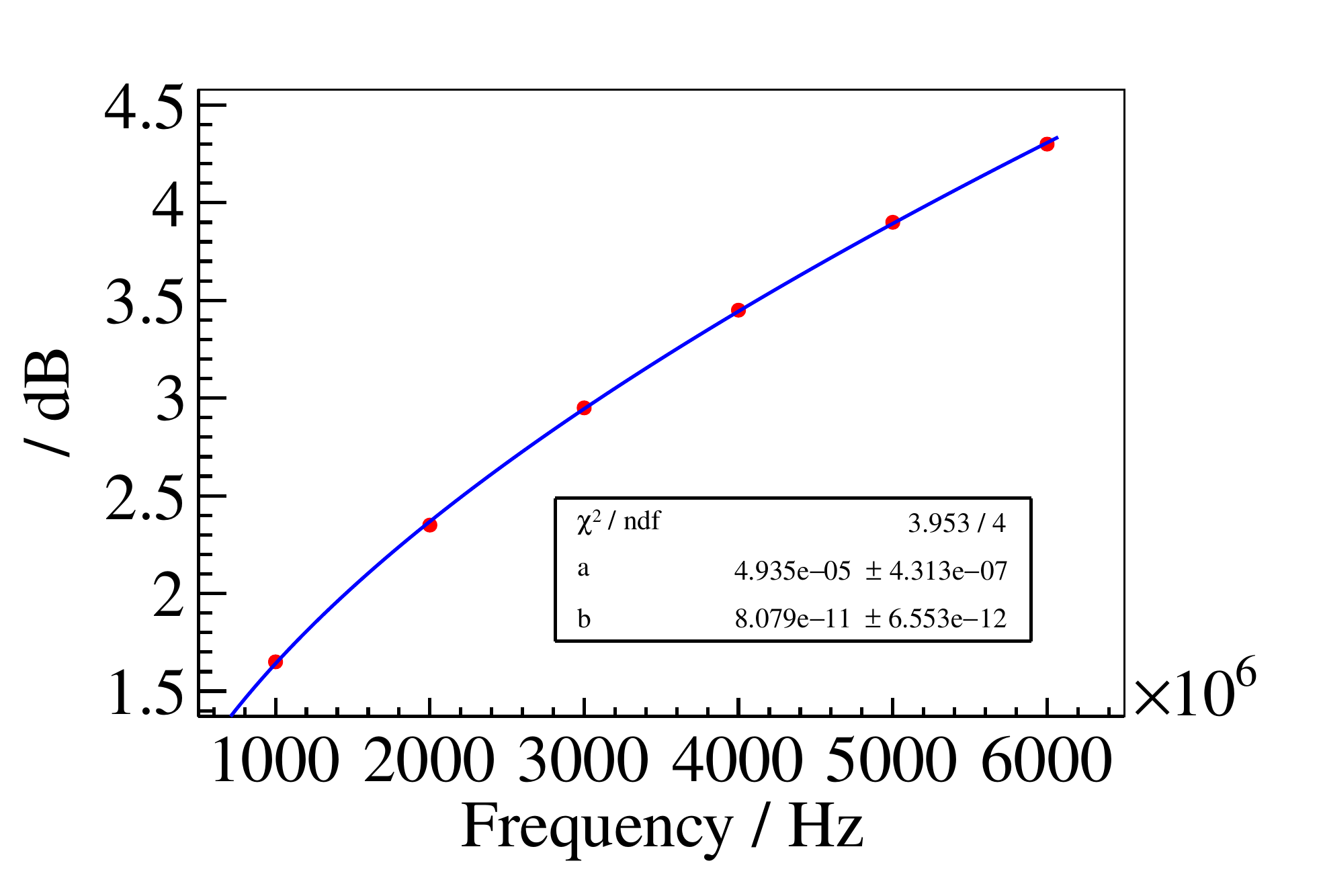} \\
    \caption{The attenuation factor curve and the fitting result of Eq.~\ref{cabFactor}. Red points are from the data sheet of the cable.}
    \label{Atte}
\end{figure}

The alternating current (AC) can only penetrate into the surface of the conductor to a certain depth (skin depth), which is known as the skin effect~\cite{SkinEffect}. The skin depth becomes smaller as the signal frequency increases, resulting in less cross-section of the conductor and larger resistance~\cite{Loss}. Therefore, the attenuation factor is a function of the signal frequency, which can be evaluated as~\cite{atte, atte1}:

\begin{equation}
    A_c=a\sqrt{f}+bf,
    \label{cabFactor}
\end{equation}

where $A_c$ is the attenuation factor expressed in dB unit and $f$ is the signal frequency. A series of $A_c$ and $f$ was obtained from the cable datasheet, which was used to determine two coefficients $a$ and $b$ through fitting (Fig.~\ref{Atte}). The attenuation factors of different frequency components with cable of $n$ meters can be calculated by $n\times A_c$.

\begin{figure}[H]
    \centering
    \subfigure[]{
        \includegraphics[width=0.21\textwidth]{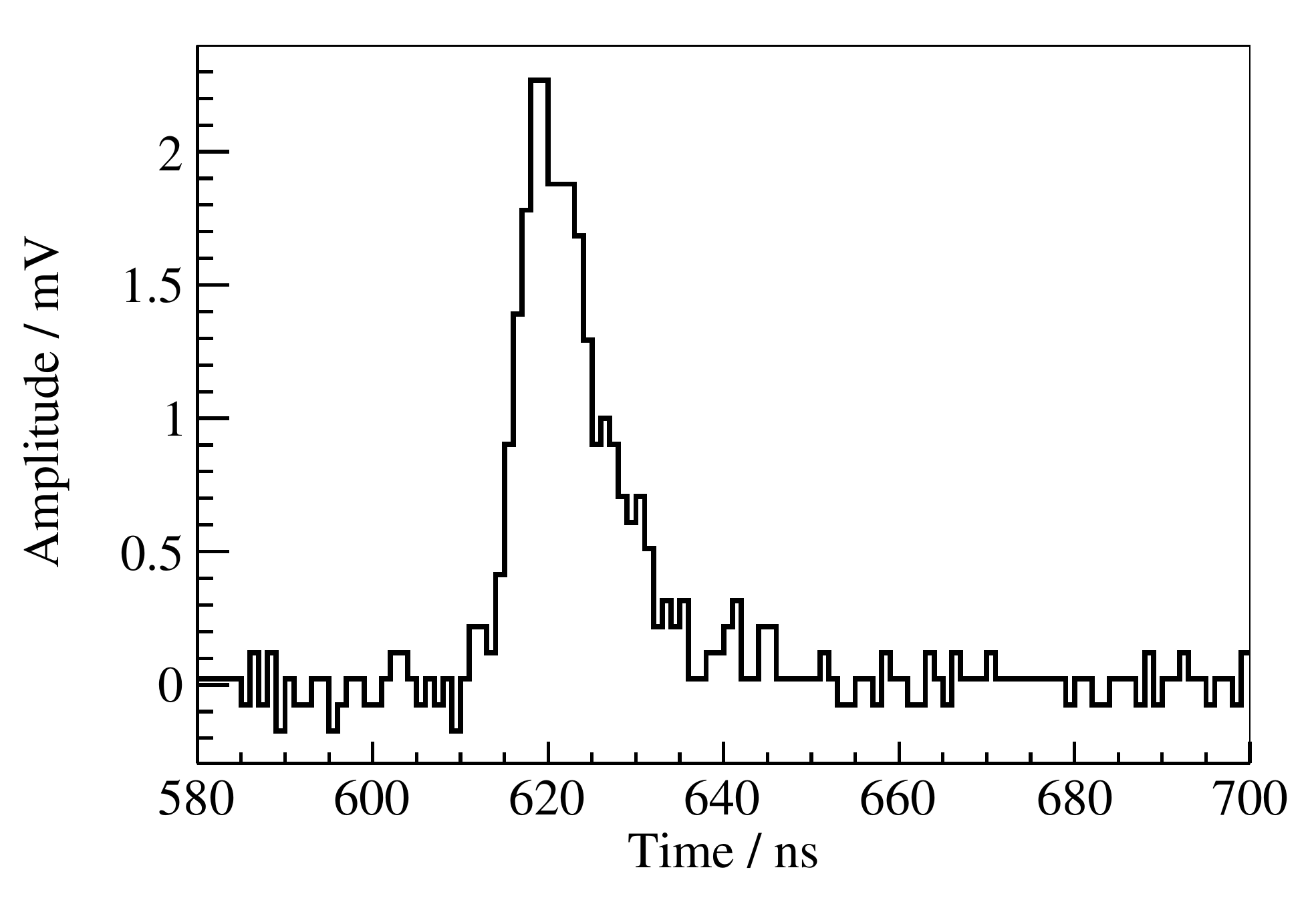}
    }
    \subfigure[]{
        \includegraphics[width=0.21\textwidth]{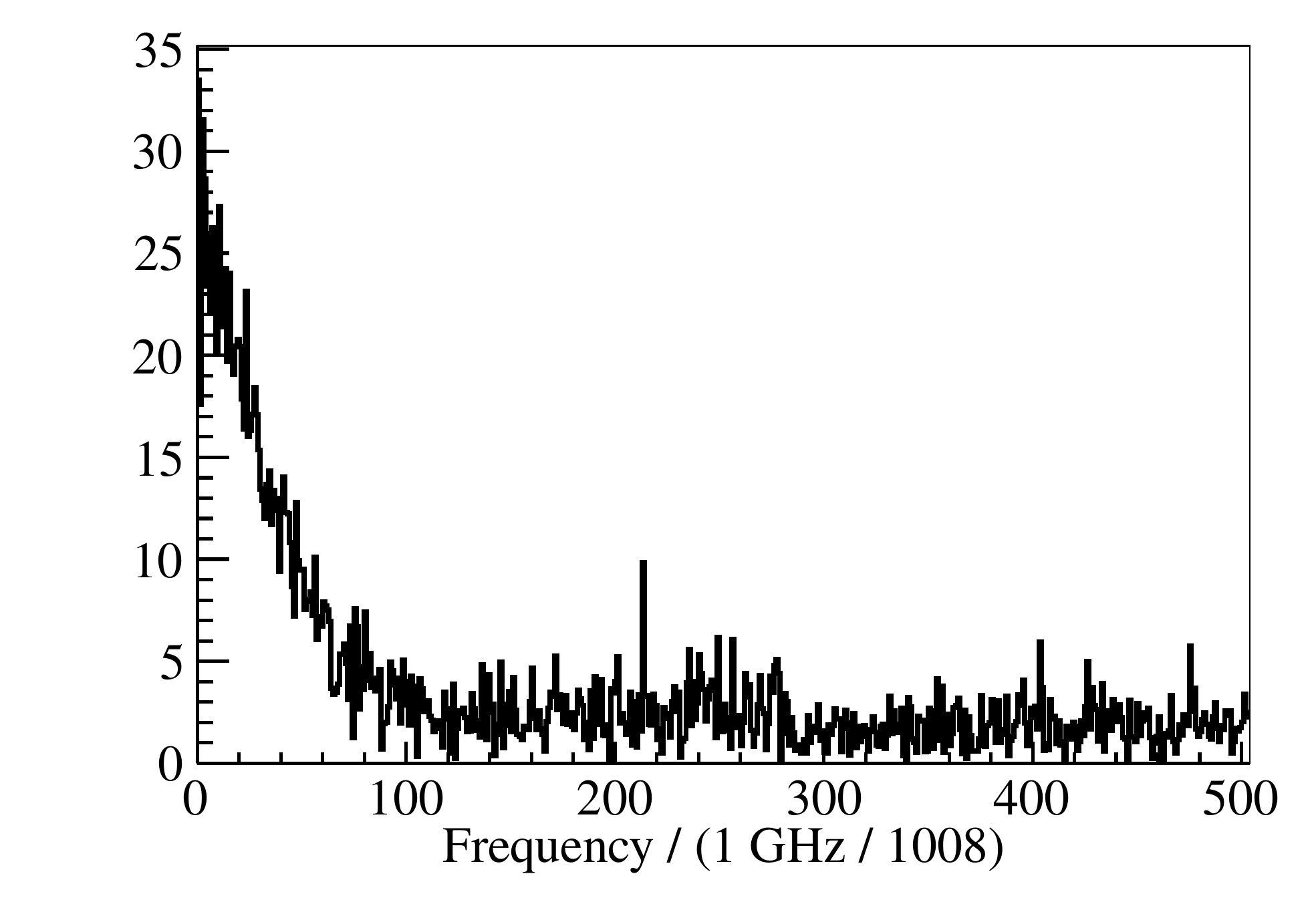}
    }
    \subfigure[]{
        \includegraphics[width=0.21\textwidth]{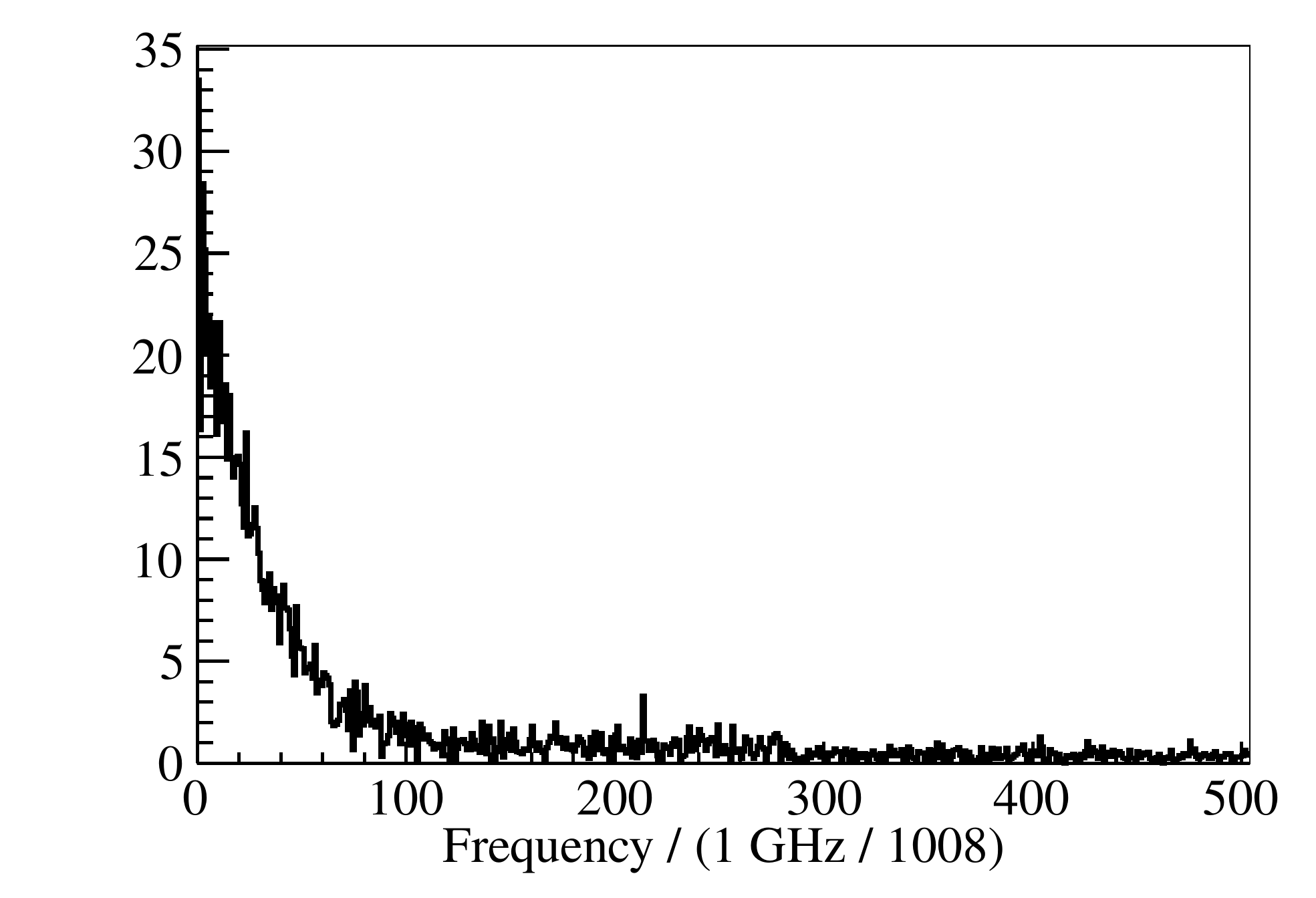}
    }
    \subfigure[]{
        \includegraphics[width=0.21\textwidth]{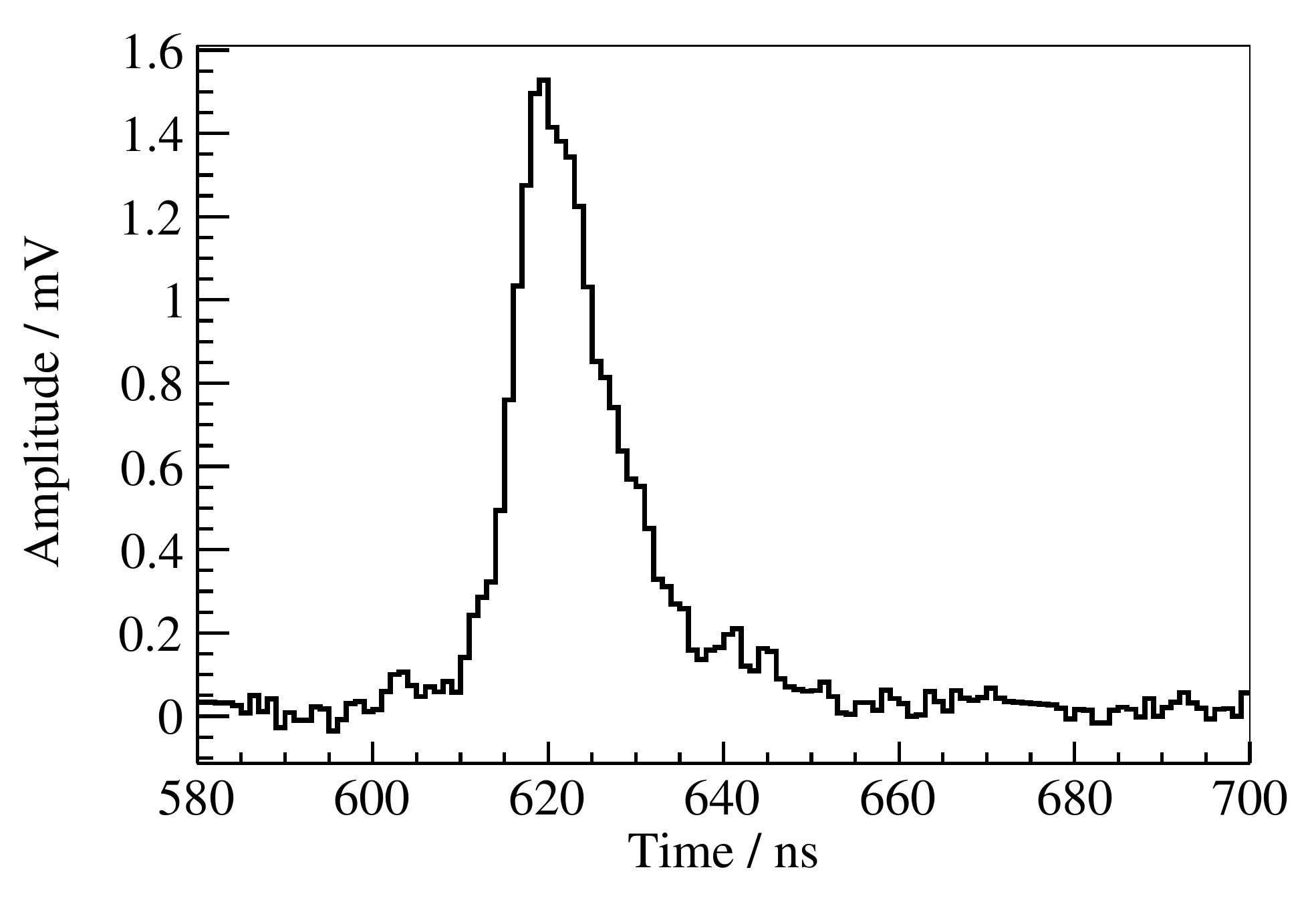}
    }
    \caption{The simulation of the attenuation process. (a) An example of the original PMT SPE waveform of 1~m cable. (b) The signal spectrum in the frequency domain. (c) The frequency spectrum after attenuation by Eq.~\ref{cabFactor}. (d) The time-domain spectrum after inverse FFT.}
    \label{AttPro}
\end{figure}

With attenuation factors of different frequencies, the PMT waveform in cases of 5~m and 10~m can be predicted based on the case of 1~m with the following steps (Fig.~\ref{AttPro}):

\begin{enumerate}
    \item Transform the signal waveform of 1~m cable (Fig.~\ref{AttPro}(a)) from the time domain into the frequency domain (Fig.~\ref{AttPro}(b)) with the Fast Fourier Transform (FFT) method.
    \item The frequency domain spectrum was multiplied by attenuation factors got from Eq.~\ref{cabFactor} and the cable length 5~m (or 10~m) (Fig.~\ref{AttPro}(c)).
    \item Inversely transforms the frequency domain spectrum back to the time domain signals (Fig.~\ref{AttPro}(d)).
    \item Get the features of attenuated waveform by fitting of log-normal function which is introduced in section \ref{section:TTS}.
    \item Repeat these steps to 2000 measured waveforms with 1~m cable, get the average of each characteristic parameter and compare with the measured results with 5~m and 10~m cable.
\end{enumerate}

\subsubsection{Measurements and comparison}

\begin{figure}
    \centering
    \includegraphics[width=0.45\textwidth]{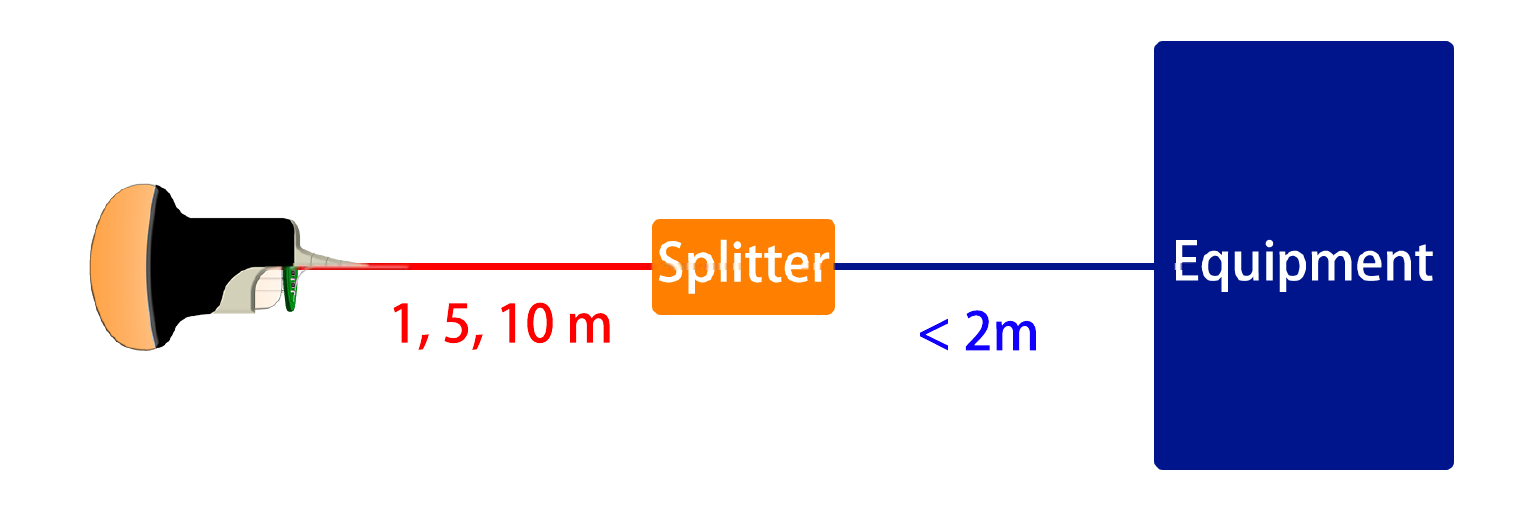}
    \caption{The cable measurement system in cartoon. Three kinds of the cables with length 1~m, the JUNO 5~m and the JUNO 10~m were measured.}
    \label{Cables}
\end{figure}

The measurement system is basically the same as that in section \ref{section:testsystem}, but a 10-times fast amplifier (CAEN N979) was added before the flash ADC (DT5751) to increase the signal-to-noise ratio. The cable from the HV divider to the splitter was the custom cable produced by AXON for JUNO and was switched between 1~m, 5~m and 10~m. The splitter and the recorder were connected by another type of cable with fixed length less than 2~m. A sketch of the measurement system is reported in Fig.~\ref{Cables} while the description of the `Equipment' part is described in Fig.~\ref{System}.

 \textbf{Gain versus cable length}
 
\begin{figure}[H]
    \centering
    \subfigure[]{
        \includegraphics[width=0.45\textwidth]{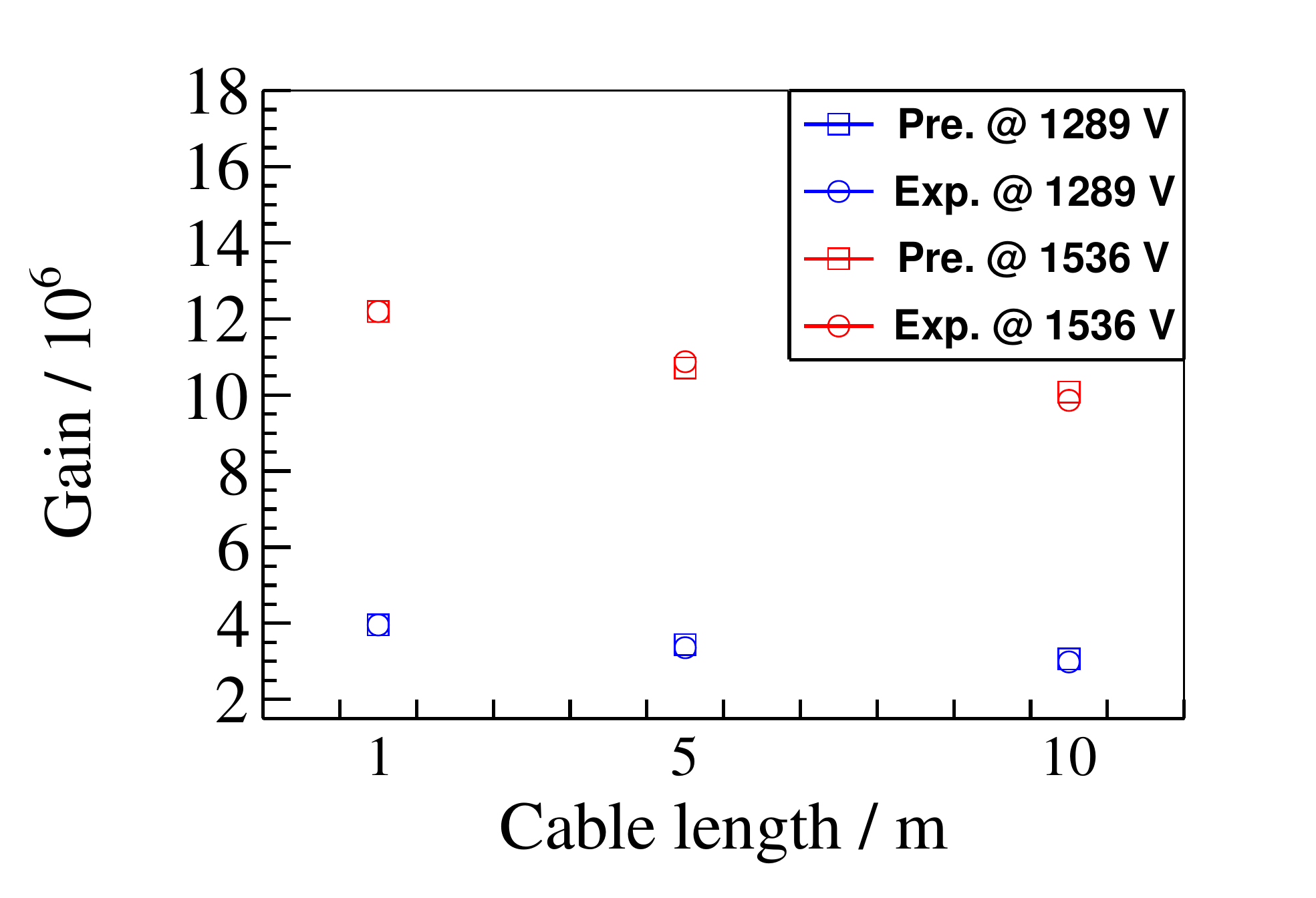}
    }
    \subfigure[]{
        \includegraphics[width=0.45\textwidth]{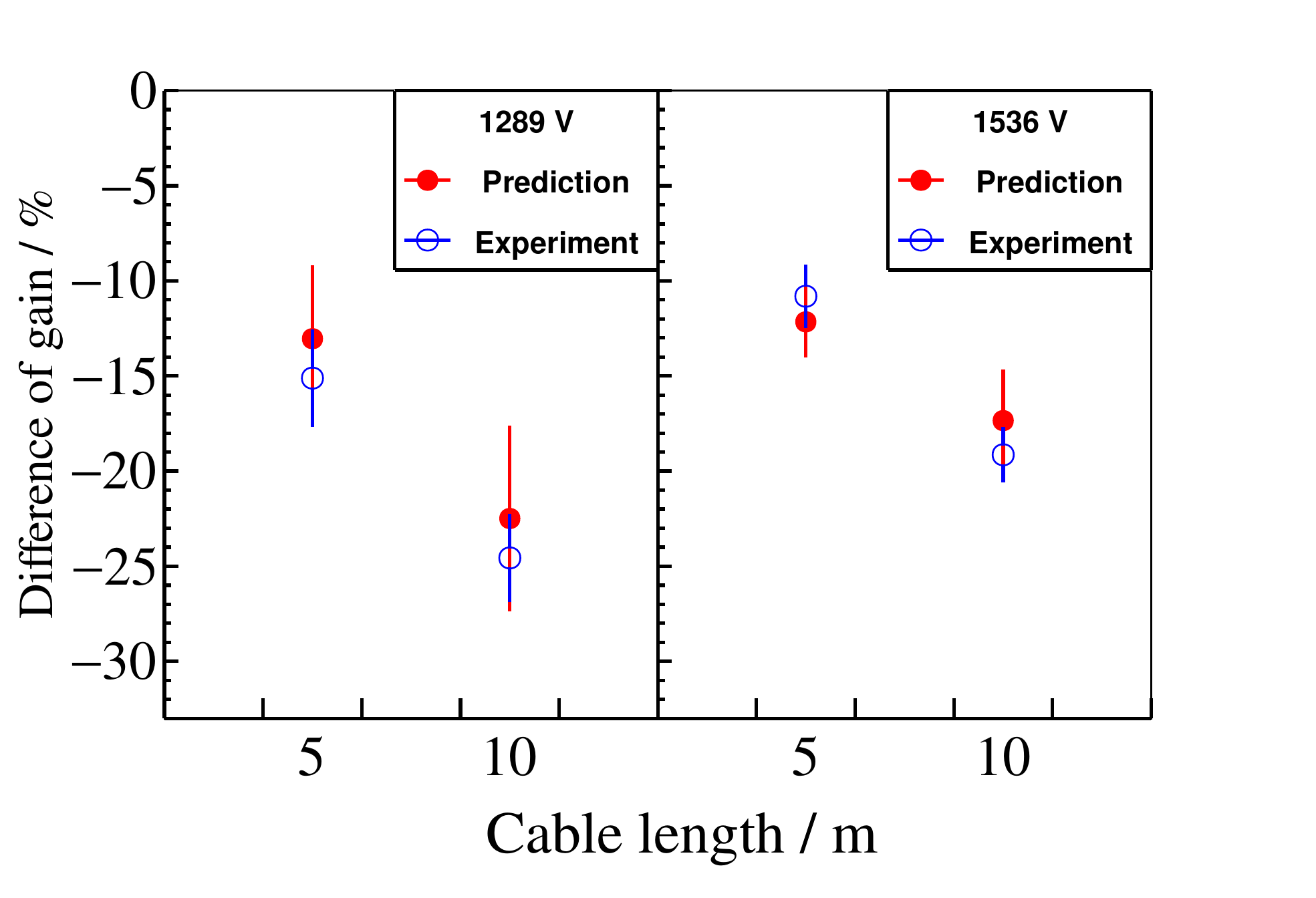}
    }
    \caption{(a) The measured gains with different cable lengths at two kinds of HV values. The 1289 V is the HV at PMT gain $3\times10^{6}$, and the 1536 V is the HV at PMT gain $1\times10^{7}$ with a 10~m cable. (b) The gain difference at 5~m and 10~m compared to the 1~m case. }
    \label{GVC}
\end{figure}

With the same calibration method presented in section~\ref{section:calibration}, PMT voltage was firstly fixed at gain $3\times10^6$ with a 10~m cable as a reference and later compared with the results obtained using other cables length. The same technique was applied to study the gain dependence at $1\times10^7$. The measured gains with 1~m, 5~m, and 10~m cable are shown in Fig.~\ref{GVC}(a). The predicted gains with cable length 5~m and 10~m calculated from cable length 1~m are also shown together. The measured and predicted gain are always in good agreement within errors for both the selected cable length (5~m, 10~m) and the voltage (1289~V, 1536~V). This proves that the prediction method is robust for what concerns the gain. There is a tendency for the gain to decrease as the cable becomes longer. The gain decreasing actually means the signal charge loss in the integration time window. The window was 100~ns in the measurement which can fully cover the SPE waveform which is shown in Fig.~\ref{TTSWave}(a) or Fig.~\ref{AttPro}(a). Compared with the gain of cable length 1~m, the measured gains of cable length 5~m and 10~m at HV 1289~V are reduced by $(13.0\pm3.9)\%$ and $(22.5\pm4.9)\%$ respectively (Fig.~\ref{GVC}(b)). Similarly, the gains at HV 1536~V are reduced by $(12.2\pm1.9)\%$ at cable length 5~m and $(17.4\pm2.7)\%$ at cable length 10~m. There is no relation between the HV and the reduction factor within uncertainties. 

  \textbf{Signal shape versus cable length}
  
The longer cable changes not only the signal charge but also the signal shape. The signal shape parameters of amplitude, rise time, fall time, FWHM were studied. All of the average values of these features were obtained from more than 2000 SPE signals. Rise time is the time from amplitude 10\% to 90\% at the rising edge. Fall time is the time from amplitude 90\% to 10\% at the falling edge. 

Just like the gain, the signals passing through a longer cable have a visible decrease in amplitude too. The amplitude decreases in the same proportion as the gain within error at the same cable and HV configuration (Tab.~\ref{ResultCable}). When the cable length increases, the reducing tendency of the signal amplitude also matches our prediction within errors.

For time-related parameters (rise time, fall time and FWHM), rise time seems not to change with different cable lengths. Fall time and FWHM become larger with a longer cable. These parameter tendencies also can be predicted by the attenuation modeling method.

The quantified measurement results of TTS and SPE resolution are also listed in Tab.~\ref{ResultCable}. TTS and SPE resolution are almost unchanged within error with the cable length changing, which is good for the choice of cable.

\begin{table}[htb]
\caption{The summary of different parameter measurements at two fixed HVs. A negative value means a decrease compared to the 1~m case. Errors of time-related features are propagated from statistical errors and system errors, and errors of other features are only from statistical errors.}
\label{ResultCable}
\begin{center}
\begin{tabular}{l|cc|cc}
\hline\hline
                 HV   &\multicolumn{2}{c}{1289 V}     &\multicolumn{2}{c}{1536 V}     \\ \cline{1-5} 
                 Cable length  & 5 m    & 10 m   & 5 m   & 10 m    \\ \hline\hline
$\Delta$ Gain (\%) & $-13.0\pm3.9$ & $-22.5\pm4.9$ & $-12.2\pm1.9$ & $-17.4\pm2.7$ \\ \hline
$\Delta$ Amplitude (\%) & $-9.0\pm1.0$  & $-19.3\pm0.8$ & $-12.0\pm1.2$ & $-22.7\pm1.0$ \\ \hline
$\Delta$ Rise time (\%)  & $-0.4\pm3.6$  & $-0.6\pm3.6$  & $-3.5\pm4.3$  & $-1.9\pm4.4$  \\ \hline
$\Delta$ Fall time (\%)  & $5.8\pm1.4$   & $20.7\pm1.5$  & $6.2\pm1.7$   & $19.8\pm1.8$  \\ \hline
$\Delta$ FWHM (\%)     & $3.4\pm1.5$   & $11.6\pm1.6$  & $1.8\pm1.8$   & $9.7\pm1.9$   \\ \hline
$\Delta$ TTS (\%) & $-6.4\pm6.7$  & $-4.0\pm6.7$  & $-1.4\pm7.9$  & $-0.7\pm8.0$  \\ \hline
$\Delta$ SPE res. (\%)    & $1.1\pm5.0$  & $1.1\pm5.5$  & $-5.8\pm3.9$  & $-5.8\pm4.2$  \\ \hline\hline
\end{tabular}
\end{center}
\end{table}

The signal shape with different cable lengths but at the same gain was studied by setting the working HV as 1244~V, 1266~V and 1289~V for the cable lengths 1~m, 5~m and 10~m, respectively. In another word, 45~V and 23~V are needed to compensate the loss of gain due to attenuation in the 5~m and 10~m long cables. The measured PMT parameters of these three cases are summarized in Tab.~\ref{SameGain}. The parameters were first calculated from each waveform, and then their averaged values were filled in the table from about 2000 waveforms.

\begin{table}[htb]
\caption{The measured parameter results of PMT with 3 kinds of cable length at the gain $3\times10^6$. Errors of time-related features include both statistical and systematic errors, and others only include statistical errors.}
\label{SameGain}
\begin{center}
\begin{tabular}{l|ccc}
\hline\hline
Gain $3\times10^6$       & 1 m               & 5 m               & 10 m              \\ \hline\hline
Amplitude (mV)  & $2.33\pm0.08$   & $2.25\pm0.08$   & $2.16\pm0.08$   \\ \hline
Rise\ time (ns) & $4.62\pm0.12$   & $4.29\pm0.12$   & $4.57\pm0.12$   \\ \hline
Fall\ time (ns) & $15.72\pm0.15$  & $15.65\pm0.16$  & $18.68\pm0.14$  \\ \hline
FWHM (ns)       & $11.78\pm0.12$  & $11.76\pm0.12$  & $12.98\pm0.12$  \\ \hline
TTS (ns)        & $1.90\pm0.09$   & $1.77\pm0.09$   & $1.74\pm0.08$   \\ \hline
SPE res. (\%)   & $44\pm2$   & $44\pm2$   & $44\pm2$   \\ \hline\hline
\end{tabular}
\end{center}
\end{table}

\begin{figure}[H]
    \centering
        \includegraphics[width=0.45\textwidth]{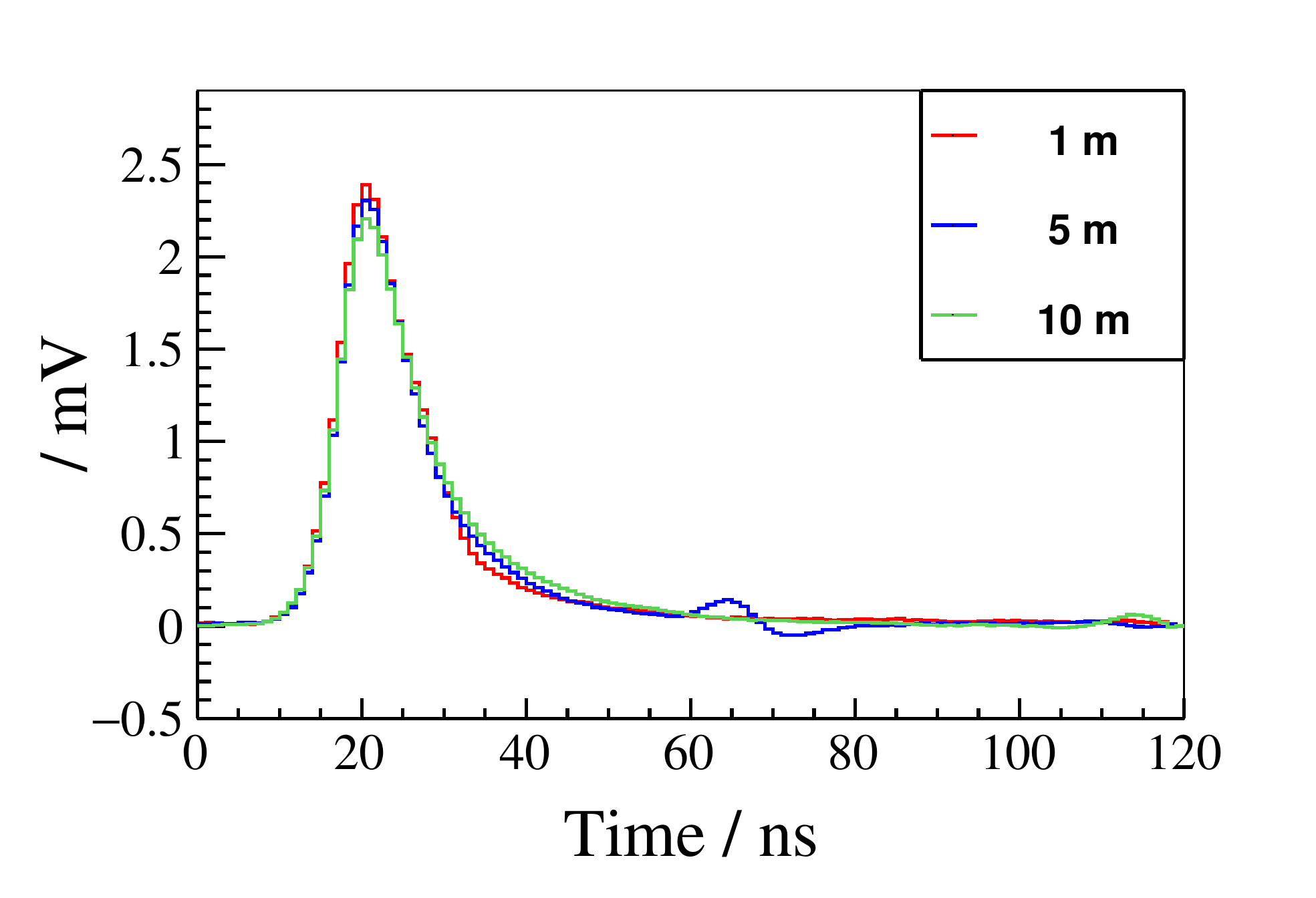}
    \caption{The average waveform with different cable lengths at gain $3\times10^{6}$.}
    \label{TVG}
\end{figure}

 Even after compensation of gain, the parameter characteristics still can not be fully compensated. Compared with the 1~m case in table \ref{SameGain}, the amplitudes of the PMT with 5~m cable and 10~m cable is reduced by about 3.4\% and 7.3\%. But compared to the result of the same HV configuration (1289~V) in table \ref{ResultCable} where the amplitudes reduce by 9\% and 19.3\%, the amplitude decreasing is about 3 times smaller after compensation. In table \ref{SameGain}, the rise time of 10~m case does not seem to change if compared with 1~m case, although there is a $\sim$~0.3~ns change for 5~m case. The fall time and FWHM of 10~m case reduce by about ($18.8\pm1.3$)\% and ($10.2\pm1.4$)\%. Both of them are not obviously changed if compared with that of ($20.7\pm1.5$)\% and ($11.6\pm1.6$)\% in Tab.~\ref{ResultCable}. It means the compensation can not obviously change the time-related parameters. The pulse with longer cable was still fatter (fall time longer and FHWM wider) and lower (amplitude lower). This can be visualized in Fig.~\ref{TVG}, average waveform with different cable lengths at gain $3\times10^{6}$. The average waveform was got by aligning the peak position fitted with Eq.~\ref{WaveFit}. The PMT waveform with cable length 10~m has a little lower amplitude and a longer tail. Thus, the gain compensation by HV increasing plays a role in counteracting the effect of cable attenuation, but the effect of cable cannot be eliminated entirely. Besides, there is a reflection in the waveform which is obvious in the 10~m case and 5~m case at the time 65~ns and 115~ns respectively. The reason of the reflection is studied in the next section.

\subsection{The signal reflection and overshoot}
\label{sec.ref}

\begin{figure}[H]
    \centering
    \subfigure[]{
    \includegraphics[width=0.35\textwidth]{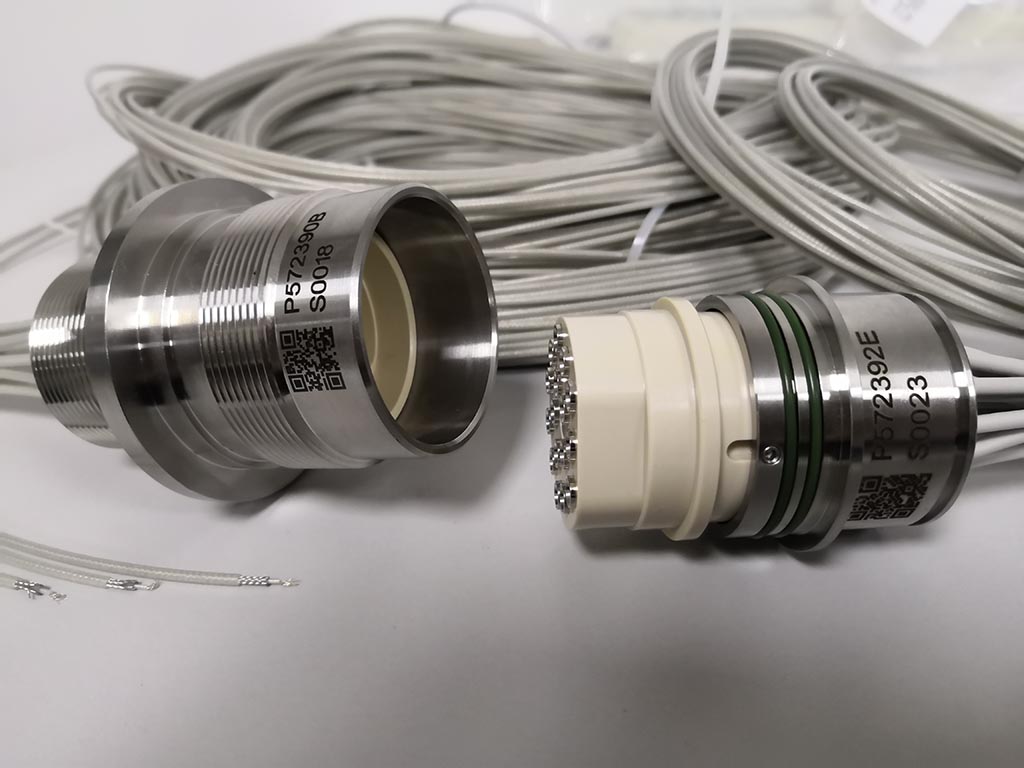}
    }
    \subfigure[]{
    \includegraphics[width=0.5\textwidth]{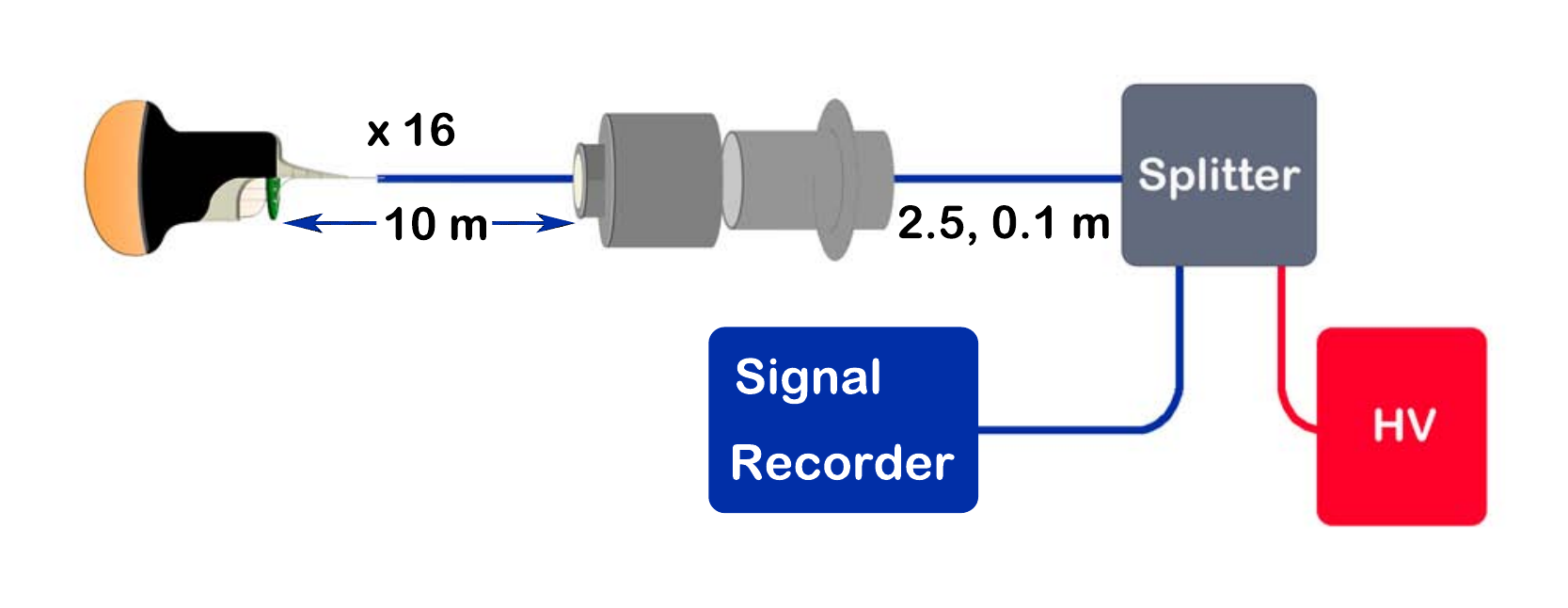}
    }
    \caption{(a) One set of a connector including one plug with 16 10-meter cables and one receptacle with 16 0.1-meter or 2.5-meter cables. (b) The schematic diagram of the reflection and overshoot measurement.}
    \label{connector}
\end{figure}

In the front-end part, there is a novel multi-channel connector for easy installation and saving cost of SPMT system. Each connector has 16 channels and can be used underwater with a depth of 40~m (Fig.~\ref{connector} (a)). If the impedance of the cable or connector is unmatched to electronics, the signal will have a certain reflection. The overshoot comes from resistor and capacitor (RC) design from splitter and PMT divider. Both signal overshoot and reflection are studied on a set of the front-end part. 
 
The connections of SPMT are shown in Fig.~\ref{connector} (b). A set of the front-end part including 16 PMTs, 16 cables, and one connector have been tested. 
The signal was amplified and was collected by a 10 bit FADC.    
Because the amplitude of overshoot is only about 1\% ($\sim$0.02~mV) of the signal peak \cite{luofj-overshoot}, which is smaller than the instrumental accuracy, a higher light density LED trigger was used to make amplitude $\sim$800 mV to get the overshoot several mV. During the measurement, two kinds of cable lengths (0.1~m and 2.5~m) were applied between the connector and splitter to confirm the reflection position. 

To precisely describe the PMT waveform with both overshoot and reflection, a new model (Eq.~\ref{Os}) was built based on the reference \cite{WaveformModel}.

\begin{equation}
    A=\frac{A_0}{exp\left(\frac{B_0-t}{\tau_0} \right)+1}exp\left(-\frac{t}{\tau_1} \right)+\frac{A_1}{exp\left(\frac{B_1-t}{\tau_2} \right)+1}exp\left(-\frac{t}{\tau_3} \right)+A_2exp\left(-\frac{(t-\mu)^2}{2\sigma^2} \right)
    \label{Os}
\end{equation}

To have a better fitting of large signals, the main pulse is rendered by the multiplication of an exponential function and a Fermi function (the first term) rather than the log-normal function used in Ref.~\cite{WaveformModel}. Same for overshoot (the second term). A Gaussian was added (the third term) to render the reflection pulse which is the new part compared with the model in Ref.~\cite{WaveformModel}. The parameters $A_0$, $A_1$, and $A_2$ are related to the amplitudes of the main pulse, overshoot, reflection, respectively. $B_0$ and $B_1$ are related to the positions of the main pulse and overshoot. $\tau_0$ and $\tau_2$ are related to the shape of the rising edge, $\tau_1$ and $\tau_3$ are related to the falling edge. $\mu$ and $\sigma$ are related to the position and width of the reflection pulse.

\begin{figure}[H]
    \centering
    \subfigure[]{\includegraphics[width=0.4\textwidth]{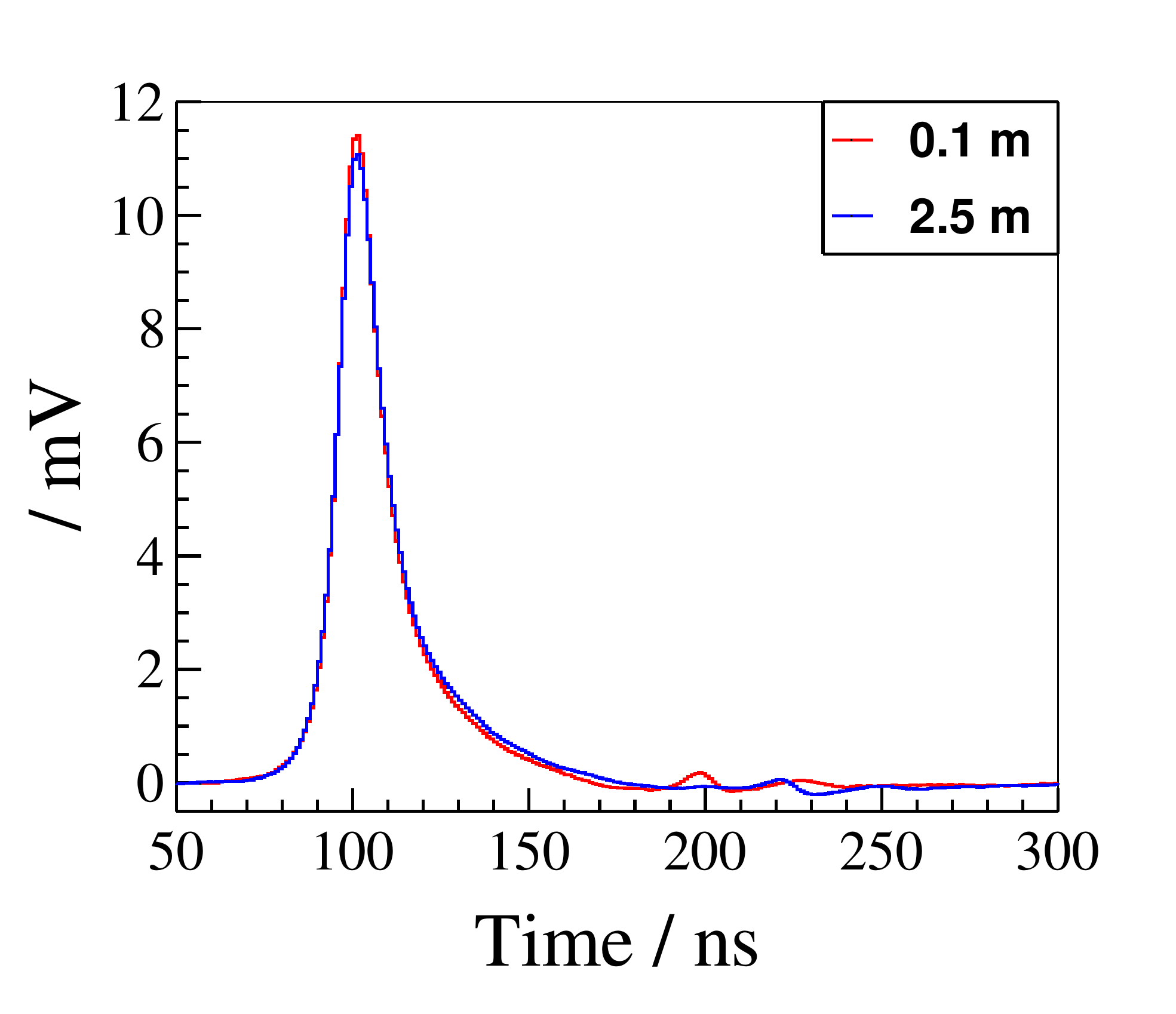}}
    \subfigure[]{\includegraphics[width=0.4\textwidth]{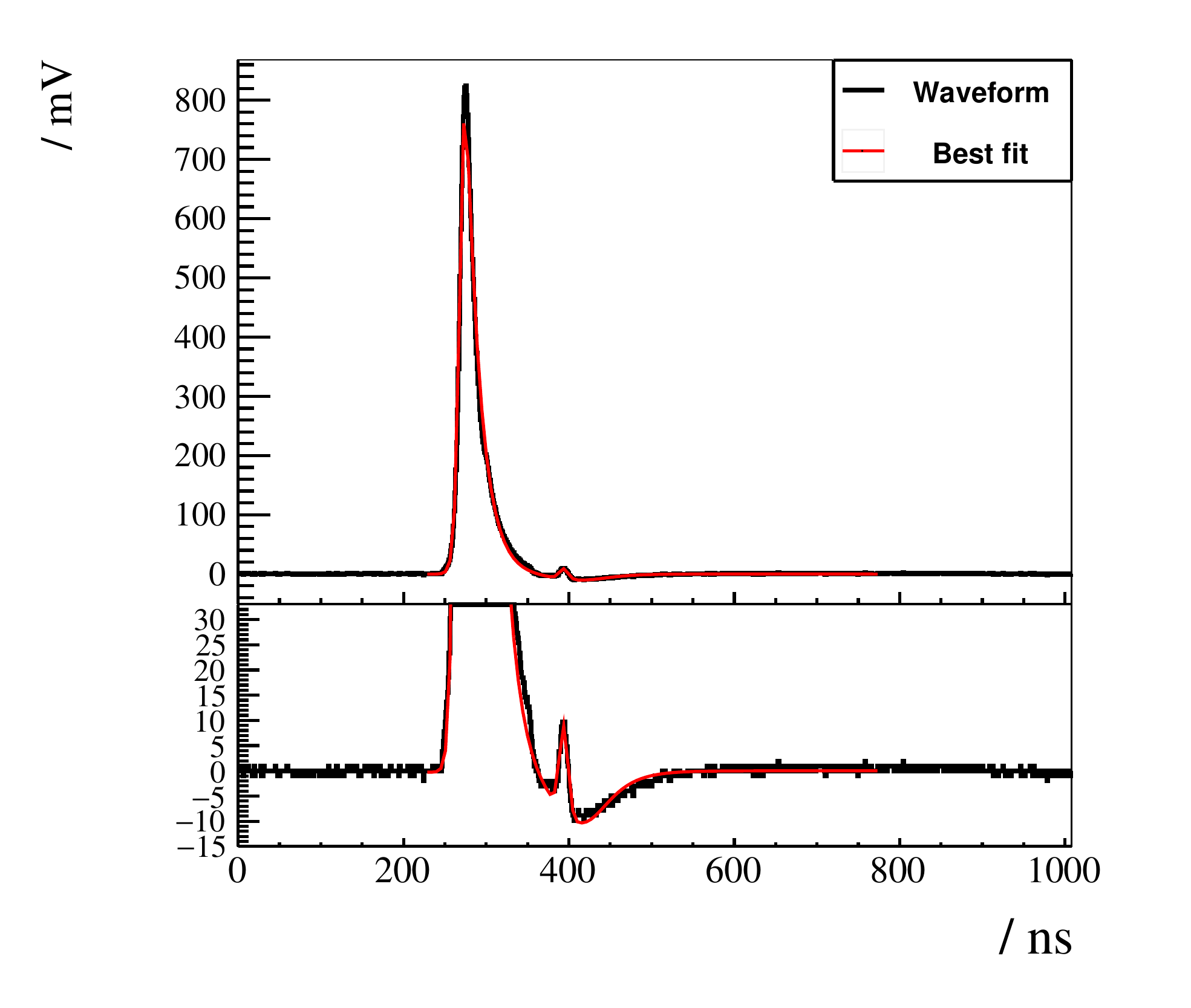}}
    \subfigure[]{\includegraphics[width=0.18\textwidth]{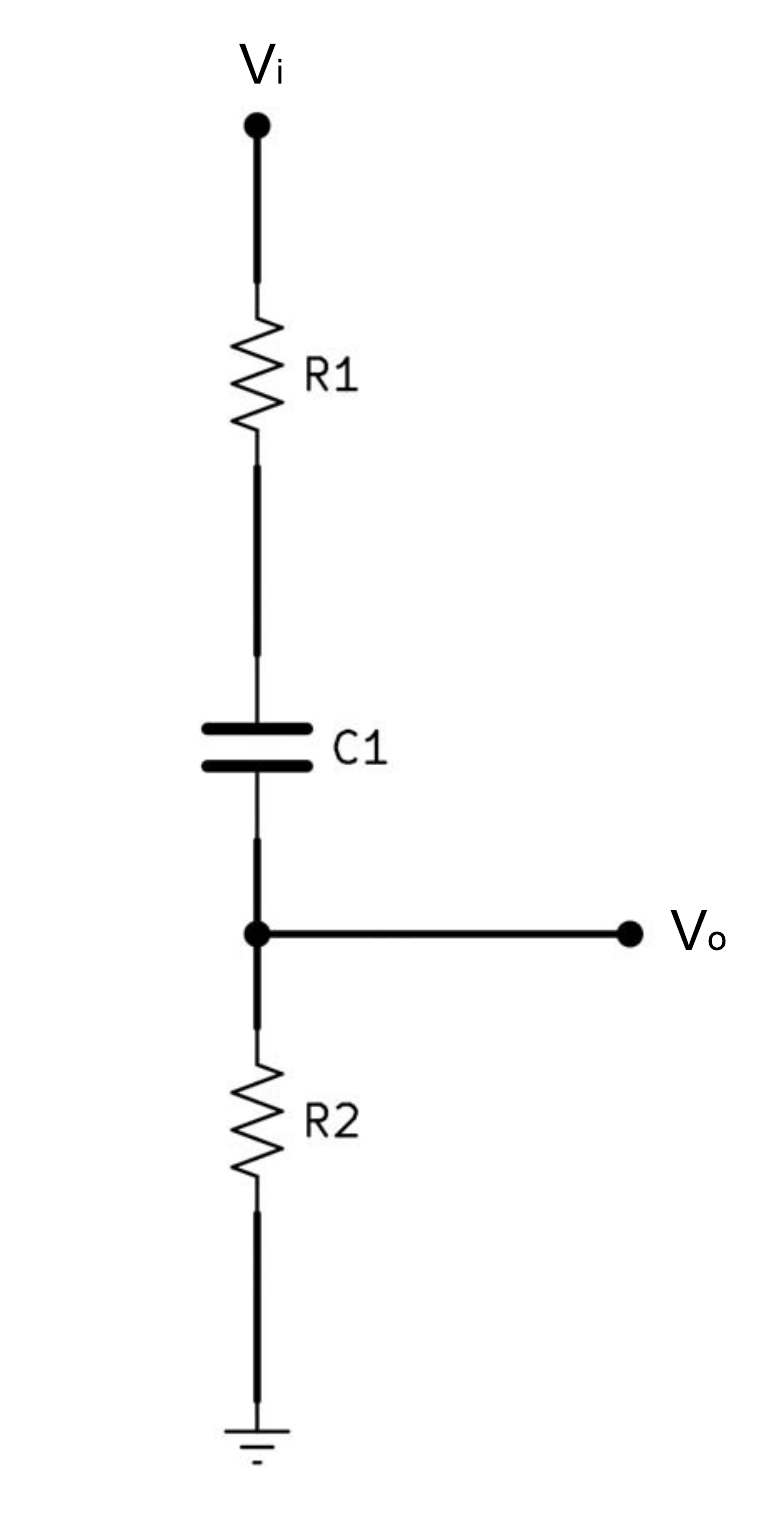}}
    \caption{(a) Average waveform collected with 0.1~m and 2.5~m cables between the connector and the splitter. (b) A large signal was collected with the 2.5~m cable and the fitting result, the overshoot and reflection area is zoomed in and shown at the bottom. (c) The simplified circuit of output part in the base and splitter.}
    \label{WaveLab}
\end{figure}

The overshoot and reflection measurement of the front-end part in the laboratory is shown in Fig.~\ref{WaveLab} (a). With the 0.1~m length cable between the connector and splitter, the time between the main pulse and reflection pulse is $(95.9\pm0.1)$~ns. With the 2.5~m length cable, it is $(118.4\pm0.1)$~ns. The reflection position changes with different cable lengths, so it should come from the splitter not from the connector. For precise calculation, the reflection time difference is about 22.5 ns which matched well with the prediction of signal transmit back and forth in the 2.4~m length cable, so the reflection position is on the splitter. The JUNO splitter will have a better design of impedance but it is beyond the study in this paper.

Overshoot is the discharge process of the capacitors in the splitter and base. Because the time constant of the capacitor in the base is larger than that in the splitter, the capacitor in the splitter mainly contributes to the overshoot. 
Based on the overshoot calculation method in Ref.~\cite{luofj-overshoot}, 
the output part of the circuit in the base and splitter is simplified (Fig.~\ref{WaveLab}~(c)) and the overshoot ratio calculation function is built. The formula is deduced from the process derived in the following.

The response of the circuit in the frequency domain can be written as Eq.~\ref{circuitResponse}, where $V_o$ is the output signal, $V_i$ is the input signal from the anode of SPMT, and $H(\omega)$ is the response function of the circuit. Concerning the R$_1$~=~1.12~$\times$~R$_2$~=~56~$\Omega$, $H(\omega)$ can be written as Eq.~\ref{responseFunction}, where $\tau=$~R$_2$~$\times$~C$_1$.

\begin{equation}
    V_{o}(\omega)=V_{i}(\omega)\times H(\omega)
    \label{circuitResponse}
\end{equation}

\begin{equation}
    H(\omega)=\frac{R_2}{R_1+R_2+1/(j\omega C_1)}=\frac{j\omega}{\left(2.12\,j\omega + 1 / \tau \right)}
    \label{responseFunction}
\end{equation}

The falling edge of the SPMT signal can be described by an exponential function and its frequency domain function can be expressed by Eq.~\ref{expon}. Where $A$ is the amplitude of a signal, $\tau_{i}$ is the time constant of the anode.

\begin{equation}
    V_{i}(\omega)=\frac{A}{\left(j\omega+1/\tau_{i}\right)}
    \label{expon}
\end{equation}

Combining the equations \ref{circuitResponse}, \ref{responseFunction}, \ref{expon}, and doing the inverse Fourier transform, the output signal in the time domain is

\begin{equation}
    V_{o}(t)=A\times \frac{1}{2.12\,(\tau_{i}-2.12\,\tau)}\times (\tau_{i}\,exp(-t/(2.12\,\tau))-2.12\,\tau\, exp(-t/\tau_{i})).
\end{equation}

Because of $2.12\times\tau\gg\tau_{i}$, the equation can be simplified as

\begin{equation}
    V_{o}(t)=A\times \frac{\tau_{i}}{2.12^2\,\tau}\,exp(-t/(2.12\,\tau)).
\end{equation}
The overshoot ratio can be expressed by $V_{o}(t)/A$, where $t$ is the time between the overshoot peak and the main pulse peak, which is 118~ns in the measurement.
The $\tau_{i}$ of the signal used in the test is $\sim16$~ns. $\tau$~=~R$_2$~$\times$~C$_1$ is 235~ns, where C$_1$ is 4.7~nF. The overshoot ratio was eventually calculated to be 1.2\%, which is very close to the measurement result of 1.1\%. The JUNO official splitter has similar capacitor and resistor, the overshoot shall be similar to this calculation.

\section{SPMT in JUNO prototype}
\label{section:prototype}

\begin{figure}[H]
    \centering
    \subfigure[]{
    \includegraphics[width=0.35\textwidth]{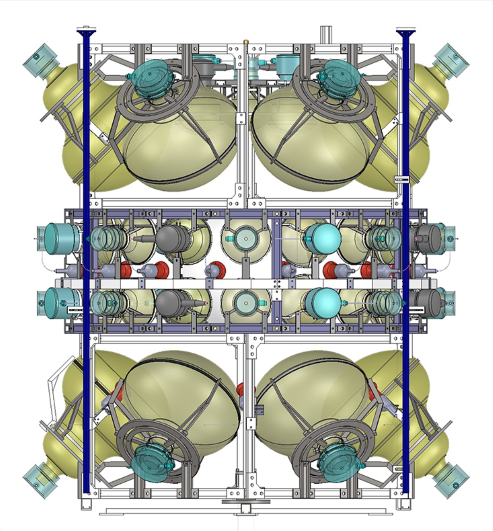}
    }
    \subfigure[]{
        \includegraphics[width=0.45\textwidth]{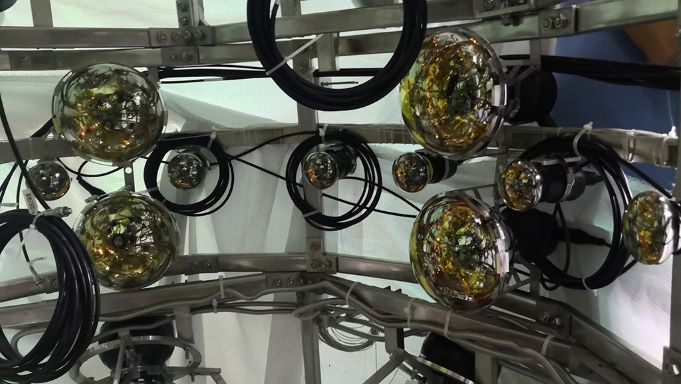}
    }
    \caption{(a) The PMT rack of JUNO prototype detector mounted with 12 20-inch PMTs, 9 8-inch PMTs and 28 3-inch PMTs. (b) The inner zoom-in view of 6 3-inch PMTs in one ring and 4 8-inch PMTs on the rack.}
    \label{Prototype}
\end{figure}

A prototype detector~\cite{zhanghq-prototype} was built by JUNO collaboration to verify the PMT performances as the first stage and then to study the cryogenic liquid scintillator performances as the second stage for the Taishan Neutrino Observatory (TAO \cite{TAOCDR}). A stainless steel tank with a size of 3~m both in diameter and in height was filled with water (first stage) or linear alkyl benzene (LAB, second stage). In the center, there is an acrylic ball with a diameter of 50~cm filled with liquid scintillator. In the second stage, a group of 16 3-inch PMTs with full instrumentation (HV divider, underwater proofing, 10~m Axon cable, and connector) were installed on the stainless steel rack surrounding the acrylic ball, together with 12 20-inch PMTs and 9 8-inch PMTs. A hand-made splitter with the same type introduced in Sec.~\ref{sec.ref} was used for each PMT to decouple HV and the signal, the latter will be recorded by FADC. The positions of these PMTs are shown in Fig.~\ref{Prototype}. The optical coverages are 0.7\%, 1.1\% and 29.9\% for 3-inch, 8-inch and 20-inch PMTs, respectively. With this setup, we were able to monitor the PMT performances in long term, at different temperatures, and to demonstrate the ability of photon detection with multiple-PMT systems.

\begin{figure}[H]
    \centering
    \includegraphics[width=0.45\textwidth]{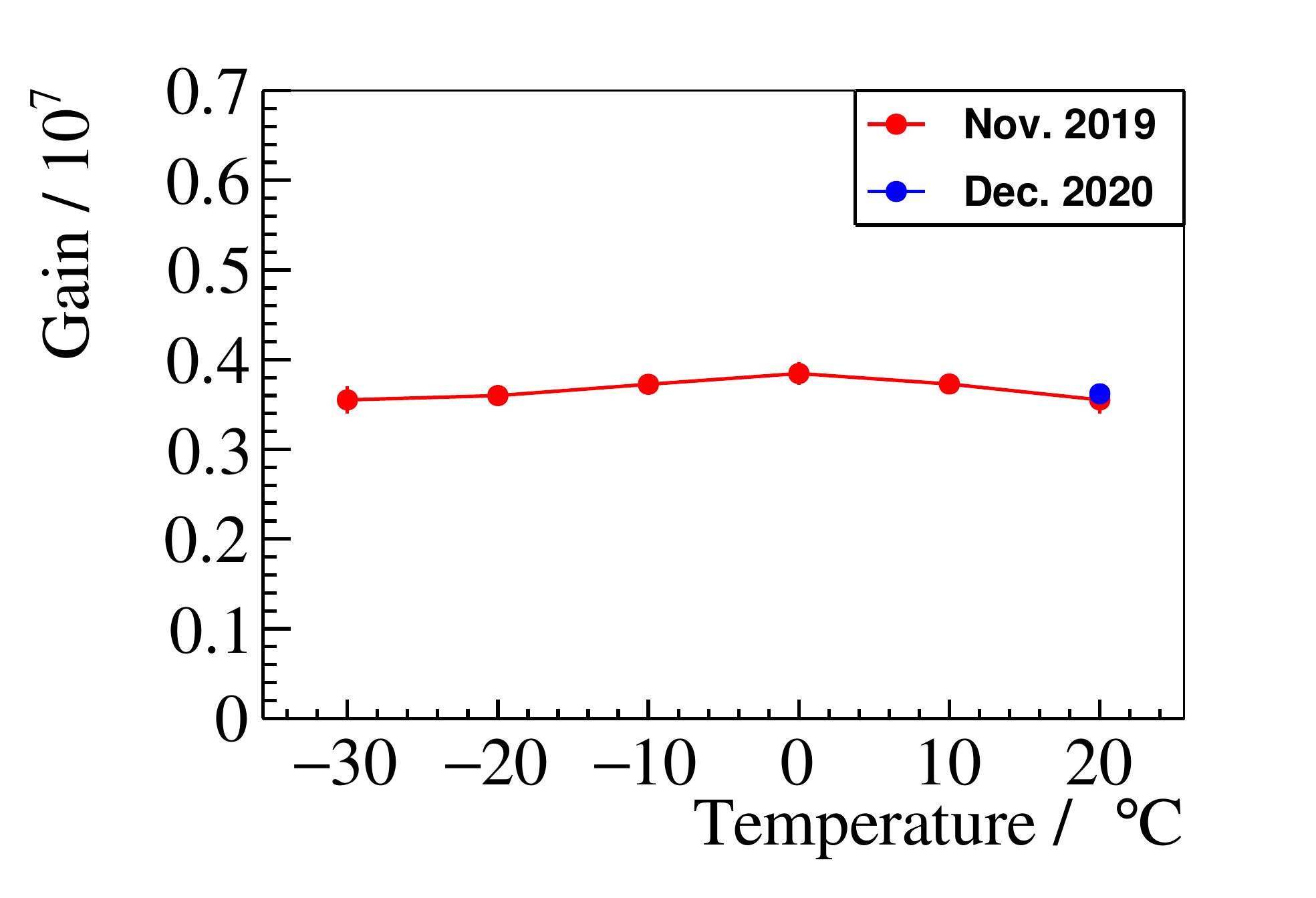}
    \caption{The average gains of 3-inch PMTs in JUNO prototype at different time or temperatures.}
    \label{GainsPro}
 \end{figure}

Starting from November 2019, a refrigeration system developed by TAO was deployed to cool down the liquid scintillator and the LAB from $20^\circ$C to $-30^\circ$C at a $10^\circ$C interval. At each temperature, the gain of each 3-inch PMT was analyzed, and their average as a function of temperature is shown in Fig.~\ref{GainsPro}. After that, the detector was back to room temperature (around $20^\circ$C), and the gain was analyzed again in December 2020, also shown in Fig.~\ref{GainsPro}. The maximum difference is $\sim$5\% compared to the result at $20^\circ$C in November 2019, showing good stability of the PMT gain, at different temperatures throughout one year. Although this kind of PMT will be operated at $21^\circ$C$\pm1^\circ$C in JUNO, this measurement provides a good reference for more possible applications.

\begin{figure}[H]
    \centering
    \includegraphics[width=0.5\textwidth]{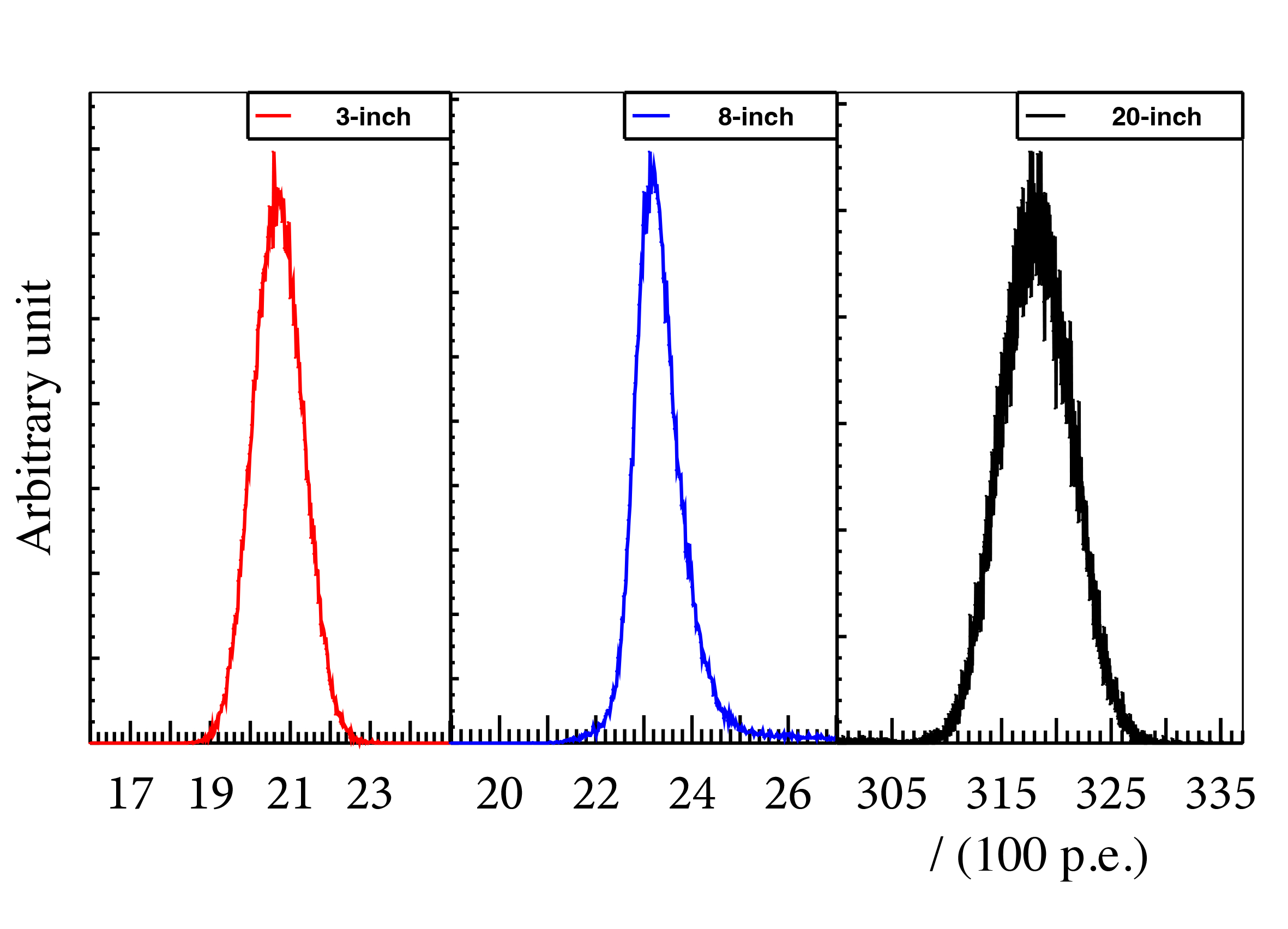}
    \caption{The photoelectron spectrum in same light intensity of 3-inch PMTs, 8-inch PMTs, 20-inch PMTs.}
    \label{ES}
\end{figure}

A LED was deployed inside the detector about 35~cm off-center to check the PMT response. Distributions of the total collected number of photoelectrons in the same trigger by 3-inch, 8-inch, and 20-inch PMTs at a certain LED light intensity are shown in Fig.~\ref{ES}. The ratio of their averages is roughly consistent with their optical coverage, with a small discrepancy coming from the position of the LED, light attenuation in the detector, and different photon detection efficiency of PMTs. These distributions were fitted with the Gaussian function and the resolutions are 2.1\%, 1.9\%, and 1.0\%. This is a good demonstration of multiple-PMT systems working in the same detector and looking at the same event. The large area PMTs provide good energy resolution and the small area PMTs extend the energy measurement to a higher range. 

Farther more, the measurements of rise time, fall time and FWHM in the JUNO prototype are $4.43\pm0.07$~ns, $18.92\pm0.39$~ns and $12.71\pm0.19$~ns, respectively. The results are consistent with those of the 10~m cable measured in laboratory (Tab. \ref{SameGain}). The reflection time is also consistent with signal theoretical back and forth transmitted time in cable. The overshoot measured in the prototype is about 2.7\% which is larger than that measured in laboratory \ref{sec.ref}. It may be enhanced by an additional reflection from read-out system.

The JUNO prototype gave a good verification of the front-end of SPMT system. 

\section{Summary}
\label{sec.summary}
The SPMT is an independent system for photon detection in the JUNO detector. It will take the task of reducing the systematic uncertainty in energy reconstruction and improving the reconstruction of muons. It consists of 25,600 3-inch PMTs, their HV dividers, waterproofing, cables, 1,600 multi-channel underwater connectors, and 200 underwater electronics boxes. In this work, we studied and confirmed a good operation of the front-end of the SPMT system including all parts except electronics boxes. PMT performances at different voltage ratios were compared using eight PMTs, and the ratio of 3:2:1 on the first 3 dynodes was chosen to be the final configuration to get better TTS with an acceptable increase of HV. The impact of the cable length was studied by recording the signal waveform of the single photoelectron. At the nominal gain of $3\times10^6$, the collected charge decreased by 13.0\% or 22.5\% with a 5~m or 10~m cable, which is consistent with the theoretical calculation. There was a 9.0\% or 19.3\% decrease of the amplitude and 5.8\% or 20.7\% increase of the fall time, however, there was almost no change of the rise time thus the TTS was not changed as well since it is mainly determined by the rising edge. To compensate for the loss of the PMT gain, 23~V or 45~V are needed. 
A small reflection was found after the main signal, and it was identified in the hand-made splitter because of a slight mismatch of impedance confirmed by adding a different length of cable between the connector and the splitter. Therefore there was no sign of reflection found in the connector. The overshoot was measured to be 1.1\% with high-intensity light, which is also consistent with the theoretical calculation.

Furthermore, the minimum unit of the front-end SPMT system including 16 channels 3-inch PMTs was installed into a prototype detector of JUNO to validate the performance after integration. The PMT gain was stable from $20^\circ$C to $-30^\circ$C in a year of operation. The total collected photoelectrons from an LED source between 3-inch, 8-inch, and 20-inch PMTs installed in the same detector were compared and it provided a good demonstration of the photon detection by multiple types of PMTs.

\backmatter

\bmhead{Acknowledgments}

We thank AXON company for the cable production. This work was supported by the National Natural Science Foundation of China No. 11975258 and 11875282, the Strategic Priority Research Program of the Chinese Academy of Sciences, Grant No. XDA10011200, the CAS Center for Excellence in Particle Physics, the Special Fund of Science and Technology Innovation Strategy of Guangdong Province.


\bibliography{sn-bibliography.bib}


\end{document}